# Verifying Compliance in Process Choreographies: Foundations, Algorithms, and Implementation

Walid Fdhila[a,∗], David Knuplesch[b], Stefanie Rinderle-Ma[c] and Manfred Reichert[d]

[a]*Security Business Austria (SBA-research), Vienna, Austria, wfdhila@sba-research.org*
[b]*alphaQuest GmbH, Ulm, Germany, d.knuplesch@alphaquest.de*
[c]*Technical University of Munich, Department of Informatics, Boltzmannstrasse 3, Garching, Germany, stefanie.rinderle-ma@tum.de*
[d]*Institute of Databases and Information Systems, Ulm University, Ulm, Germany, manfred.reichert@uni-ulm.de*

ABSTRACT

The current push towards interoperability drives companies to collaborate through process choreographies. At the same time, they face a jungle of continuously changing regulations, e.g., due to the pandemic and developments such as the BREXIT, which strongly affect cross-organizational collaborations. Think of, for example, supply chains spanning several countries with different and maybe even conflicting COVID19 travelling restrictions. Hence, providing automatic compliance verification in process choreographies is crucial for any cross-organizational business process. A particular challenge concerns the restricted visibility of the partner processes at the presence of global compliance rules (GCR), i.e., rules that span across the process of several partners. This work deals with the question how to verify global compliance if affected tasks are not fully visible. Our idea is to decompose GCRs into so called assertions that can be checked by each affected partner whereby the decomposition is both correct and lossless. The algorithm exploits transitivity properties of the underlying rule specification, and its correctness and complexity are proven, considering advanced aspects such as loops. The algorithm is implemented in a proof-of-concept prototype, including a model checker for verifying compliance. The applicability of the approach is further demonstrated on a real-world manufacturing use case.

## 1. Introduction

Gartner regards interoperability as *"strategic imperative"*[1] for healthcare. Especially the global push by digitalization and the current pandemic require the collaboration and integration of (business) partners and organizations. Process technology serves as enabler for process-oriented collaborations between distributed business partners, realized and implemented through so-called *process choreographies*. Applications include healthcare (34), blockchain-based processes (47; 46), multi-modal logistics scenarios (7; 10), and supply chains (22).

Digitalization and ongoing changes due to, for example, the pandemic situation or the BREXIT flood enterprises and organizations with updated or even new regulations at a fast pace. For example, *"bank regulations change every 12 minutes"*.[2] Regulatory frameworks comprise application-independent frameworks such as the GDPR on *"data processing boundaries of the personal data of European Union's citizens"* (65) and the ISO 27001 security standard[3] as well as application-specific ones, e.g., the WHO regulations defined in the context of COVID19[4]. As a consequence, in our globalized world, regulations and their changes affect process collaborations (54) and lead to an increased need for *compliance verification* in process choreographies.

### 1.1. Problem Statement

What are the particular challenges with respect to compliance verification in process choreographies? Let us illustrate them by an example. Figure 1 depicts the choreography model of a supply chain scenario adapted from (22). It involves five process partners, i.e., *Bulk Buyer*, *Manufacturer*, *Middleman*, *Supplier*, and *Special Carrier* that interact through message exchanges. First, the *Bulk Buyer* orders a set of products from the *Manufacturer* (e.g., an aircraft). The manufacturing of the product requires several sub-products (intermediates) to be provided by different suppliers. In this scenario, we assume that only one intermediate is required and provided by the *Supplier*. After processing the order, hence, the *Manufacturer* sends an order for the intermediate (e.g., the fuselage or engines) to the *Middleman*. The *Middleman* forwards the order of the intermediate to the *Supplier* and sends an order for a special transport to the *Special Carrier*. The *Special Carrier* requests the details on the transport from the *Supplier* and the *Supplier* provides them to the *Special Carrier*, followed by sending the waybill for the intermediate. The *Special Carrier* sends a notice on the arrival of the intermediate to the *Manufacturer*, which then delivers the product to the *Bulk Buyer*.

---
∗Corresponding author
✉ wfdhila@sba-research.org (W. Fdhila)
ORCID(s): 0000-0002-5245-6128 (W. Fdhila); 0000-0001-5656-6108 (S. Rinderle-Ma); 0000-0003-2536-4153 (M. Reichert)
[1]https://gtnr.it/3vGFB7f
[2]https://thefinanser.com/2017/01/bank-regulations-change-every-12-minutes.html/
[3]https://www.iso.org/isoiec-27001-information-security.html
[4]https://www.who.int/teams/regulation-prequalification/covid-19

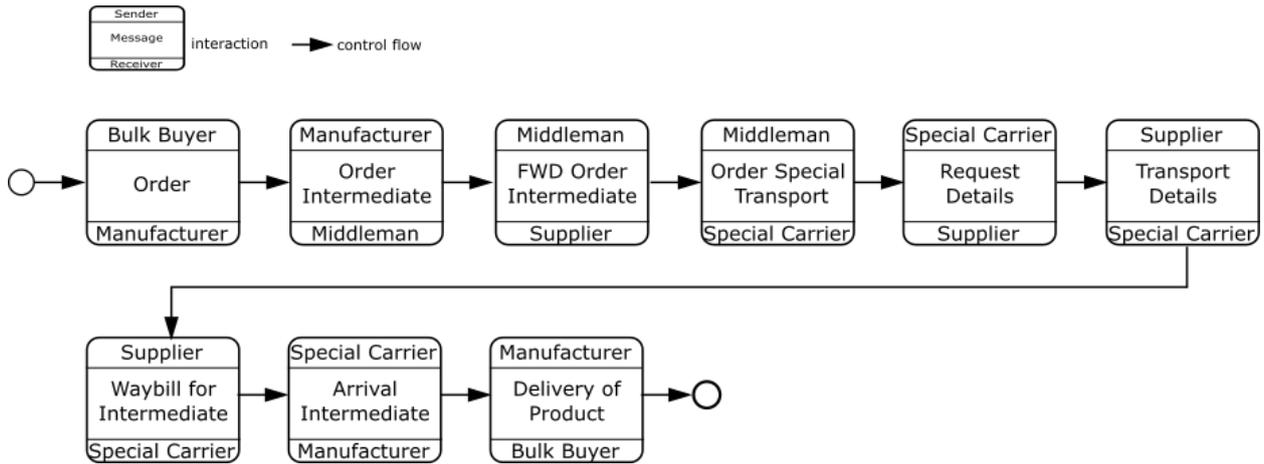

**Figure 1:** Choreography model for a supply chain – running example with five process partners

Imagine now that the partners and the choreography are subject to the *Global Compliance Rules (GCR)* depicted in Fig. 2, which stem from legal regulations and standards such as GDPR or ISO 27001:

C1 After *Production* a *Final test* must be performed.
C2 *Pack Intermediate* is required before *Transport Intermediate*.
C3 Each *Transport intermediate* requires *Permission of authority*. Further on, the transporter must pass a *Safety Check*.

Obviously, none of the GCR can be directly verified on basis of the choreography model in Fig. 1 as none of the public and message exchanging tasks corresponds to any of the tasks referred to in the GCR.

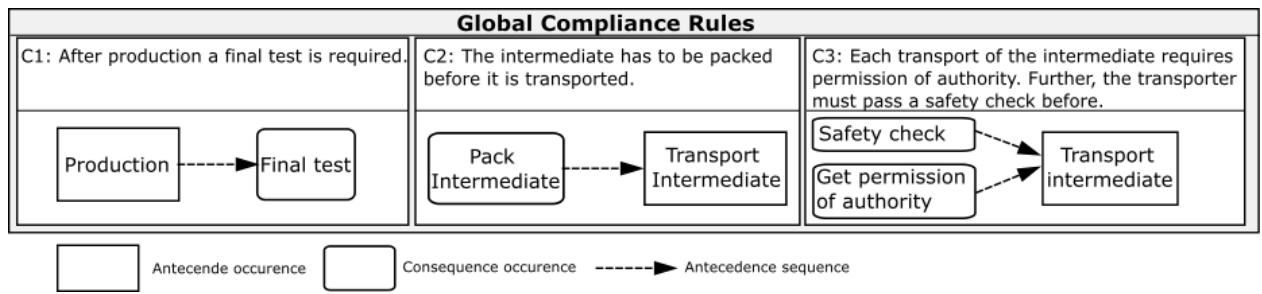

**Figure 2:** Global Compliance Rules imposed on supply chain choreography

Let us have a look at the *public processes* of the partners involved in the choreography as shown in Fig. 3. These public process models contain all public tasks that are visible to the other partners, including the tasks that exchange messages, but also other visible tasks such as *Production* at the *Manufacturer*. Based on the public process models, C1 and C2, as depicted in Fig. 2, can be verified: C1 refers to public tasks of the *Manufacturer* process, which obviously complies with C1, i.e., public task *Production* is followed by public task *Final Test*. C2 can be verified over the *Supplier* and *Special Carrier* processes. The order between tasks *Pack Intermediate* and *Transport Intermediate* is determined by the message exchange between sending and receiving *Waybill Intermediate*. As opposed to C1 and C2, C3 cannot be verified based on the public processes of the partners as there are no public tasks for *Safety check* and *Get permission of authority*.

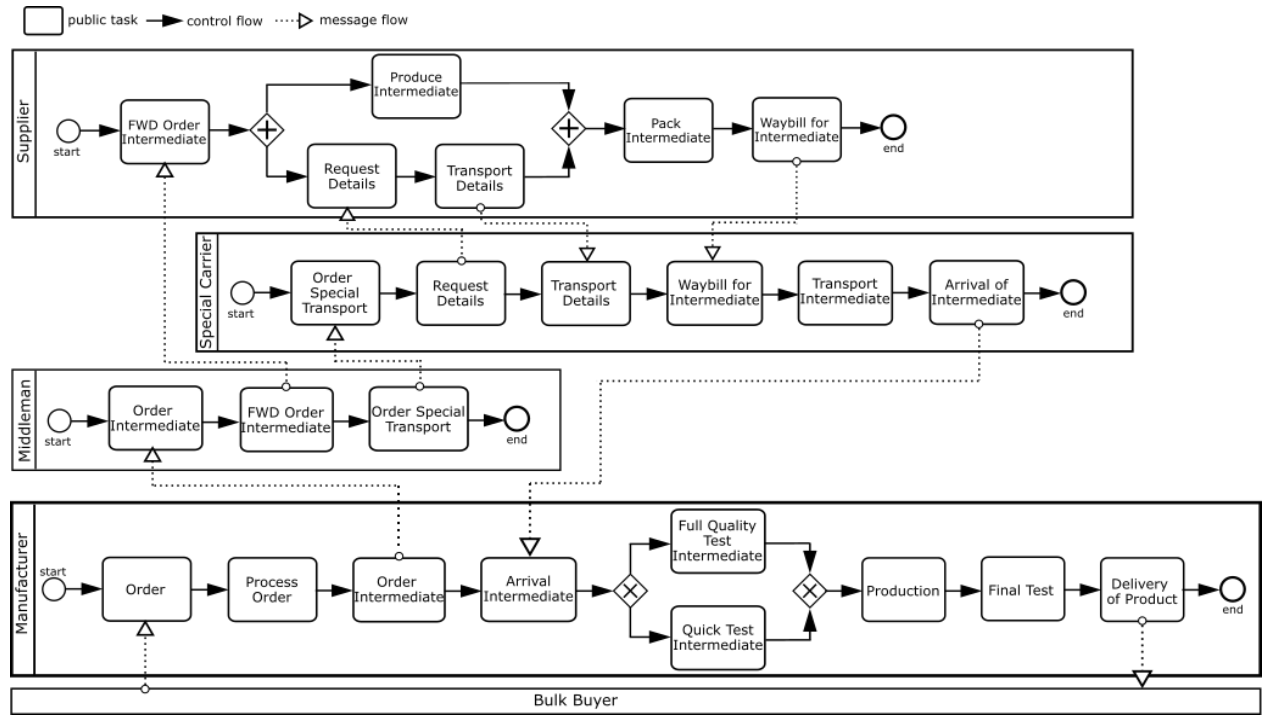

**Figure 3:** Public processes (collaboration model) – running example with five process partners cooperating in a supply chain (adapted from (22))

The presumption is that C3 also refers to *private tasks* of the partners, i.e., tasks that are only present in the *private process models* of the partners. In general, private process models of the partners implement and possibly extend the behavior of the corresponding public models. As opposed to public tasks, private tasks are not visible to the partners. Figure 4 shows the private process models of partners *Special Carrier* and *Middleman* where *private tasks* are highlighted in gray color. Although private tasks are usually hidden to other partners, restrictions over them might exist. In the supply chain, for example, C3 refers to private tasks *Safety Check* for partner *Special Carrier* and *Get permission of authority* for partner *Middleman*. If private tasks are affected by a GCR, no information about how and when these tasks are executed, or how they are connected to other nodes of the corresponding private process model, becomes visible to the other partners. Usually, this happens when a collaborating partner p1 imposes the execution of a specific task that must exist in its private process and comply with a given rule involving another partner p2. Partner p1 should then assure the existence of such task and that it follows the imposed rule.

As can be seen from the example depicted in Figs. 1–4, GCRs constrain actions of multiple partners and/or the interactions between them. Ensuring the compliance of process choreographies with a GCR is crucial and challenging (32) as a partner *"only has the visibility of the portion of the process under its direct control"* (48). Reconsider GCR C3 as an example. It asks for a *safety check* accomplished by a private task of the *Special Carrier*. To cope with this issue, *assertions* can be used. An assertion (A) corresponds to a commitment of a partner guaranteeing that its private/public process complies with the imposed rule (22). Figure 5 depicts the two assertions A1 and A2: the *Middleman* agrees to get the permission of the authority before ordering the special transport (A1). Moreover, the *Special Carrier* commits to perform a safety check before transporting the intermediate (A2). In combination, assertions A1 and A2 enable

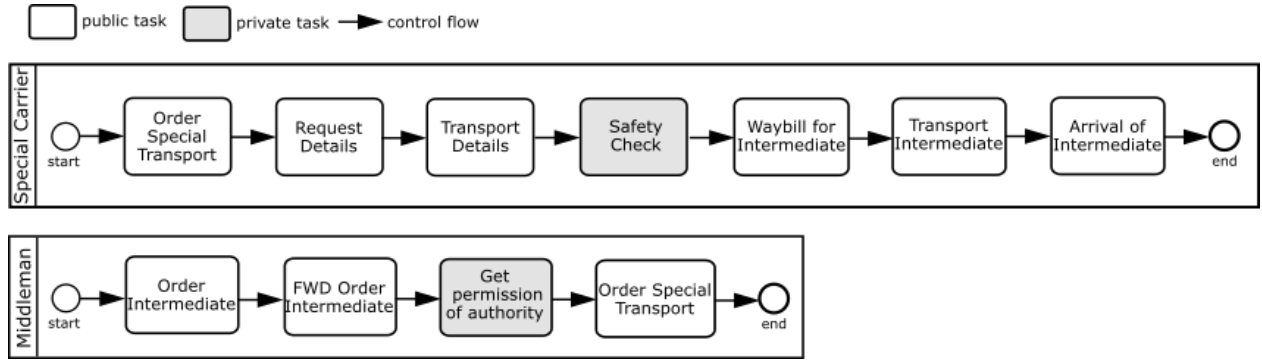

Figure 4: Private processes of partners *Special Carrier* and *Middleman*, omitting message exchanges (adapted from (22))

checking GCR C3.

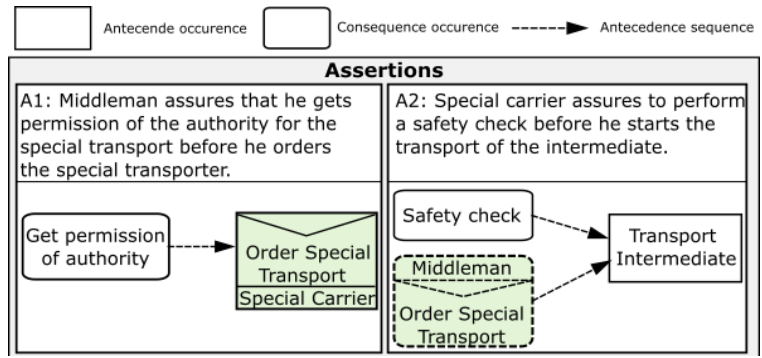

Figure 5: Assertions by partners *Middleman* and *Special Carrier*

### 1.2. Contribution
Overall, this leads to the overarching research question *RQ* tackled in this work:

> *RQ: How to verify GCRs in a decentralized setting of a process choreography where no central coordinator with complete knowledge on the private and public tasks of all partners exists?*

In literature, there is a *"knowledge gap"* when it comes to compliance verification in process choreographies (32). (48) tackles the problem of checking a GCR on private tasks based on IoT-enabled artifacts. However, not all process choreography settings with compliance requirements feature IoT-enabled artifacts. Hence, this works aims at providing a formal approach that is independent of any technology or application. The central idea is to decompose the GCR into assertions in a lossless way, i.e., the verification of all assertions guarantees the one of the GCR. Note that this solves the problem as assertions may be checked separately by each of the partners. Hence, infringing the privacy of any partner is avoided.

The decomposition algorithm presented in this article exploits transitivity properties of the underlying GCR specification and was originally presented in (23). The decomposition relies on transitivity properties of the underlying GCR specification. The transitivity properties are shown by the example of a translation to first order predicate logic. Similarly, for example, (57) presents declarative patterns based on Linear Temporal Logic (LTL).

In our approach, GCRs are specified in a pattern-based and visual way using the eCRG formalism (40). This means that a GCR may be composed of so called *antecedence patterns* and *consequence patterns*. The patterns can be connected reflecting pre-/post-conditions of the respective GCR. C1 in Fig. 2, for example, connects antecedence pattern *Production* with consequence pattern *Final test*, demanding that after the production a final test is required. Note that antecedence and consequence patterns may require *occurrence* (i.e., something must happen) and *absence* (i.e., something must not happen). In (23), we relied on simple rules that consist of single antecedent and multiple occurrence patterns. Aside the decomposition algorithm itself, (23) provides basic proofs, simple GCR decomposition scenarios, and the embedding of the approach in the overall digitalized change and compliance management framework $C^3$Pro.[5] This article extends and elaborates the results presented in (23) in several directions:

- We allow for additional and more complex compliance rules with multiple antecedence patterns. This significantly increases the complexity of the theoretical considerations as well as the one of the provided GCR decomposition scenarios. As a result, we obtain new theorems and algorithms.

- The decomposition proofs are extended to cover the additional complexity of the GCR; in particular they now consider loops as well.

- The decomposition algorithm with extensions is prototypically implemented and integrated with the $C^3$Pro framework, which deals with both change and compliance in process choreographies.

- A model checker for verifying decomposition correctness is provided.

- A manufacturing use case illustrates the applicability of the approach. Specifically, the use case demonstrates the applicability of the approach beyond regulatory compliance, i.e., it shows how decomposition can be used to lift implicit connections to explicit assertions.

- The related work section is significantly extended.

The remainder of the paper is structured as follows: Section 2 provides fundamentals required for understanding this work, and Section 3 introduces the foundations for GCR decomposition (including transitivity theorems). Section 4 then presents the decomposition algorithm for global compliance rules, whereas Section 5 deals with the automated verification of the resulting decompositions based on model checking. Sections 6 and 7 cover the evaluation of the approach, i.e., the implementation and application of the algorithms. Section 8 discusses related work. Section 9 concludes the paper with a summary and an outlook.

## 2. Fundamentals

This section presents definitions and formal backgrounds for global compliance rules (GCRs) to be obeyed by a process choreography $y$. A *choreography* includes three types of overlapping models: (i) a *private model* representing the **executable** process and including both private and public activities (see Fig. 4 for examples of private process models), (ii) a *public model* (also called the *interface* of the process) that solely includes the public activities and the interactions of a given partner (see Fig. 3 for the public process models of our running example), and (iii) a *choreography model* providing a global view on the interactions between all partners (see Fig. 1 for the choreography model of our running example) (21). Compared to (21; 2), this paper assumes that public activities are not necessarily interactions with other partners, but may additionally represent tasks made visible by the corresponding partners. Therefore, both interactions and non-interaction public activities of a single partner are described in a public model. The latter serves as public (restricted) view on the private model of the partner, which *"describes the internal logic of a partner including its private and public activities"* (6). For a formal definition of process choreography, we refer to Definition 1.

**Definition 1 (Choreography (21)).** *We define a choreography y as a tuple*
*($\mathcal{P}$, $\mathcal{G}$, $\Pi$, $\mathcal{L}$, $\psi$, $\Gamma$, $\xi$) where*

- $\mathcal{P}$ *is the set of all participating partners.*
- $\mathcal{G}$ *is the choreography model representing the interactions $\mathcal{I}$ between partners in $\mathcal{P}$ (cf. Fig. 1).*
- $\Pi = \{\pi_p\}_{p \in \mathcal{P}}$ *is the set of all private models (cf. Fig. 4).*
- $\mathcal{L} = \{l_p\}_{p \in \mathcal{P}}$ *is the set of all public models (cf. Fig. 3).*
- $\psi = \{\psi_p : l_p \leftrightarrow \pi_p\}_{p \in \mathcal{P}}$ *is a partial mapping function between nodes of the public and private models.*
- $\Gamma: l \leftrightarrow l'$ *is a partial mapping function between nodes of different public models.*
- $\xi : \mathcal{G} \leftrightarrow l \times l$ *is a partial mapping function between nodes of the choreography model and the public models.*

---

[5]<http://www.wst.univie.ac.at/communities/c3pro>

As depicted in Figures 1, 3 and 4, choreography, public and private models are defined as graphs, where nodes are either activities (i.e., interaction, public or private activities) or gateways (e.g., sequence, exclusive or parallel), and arrows are the dependencies between them. As described above, each of these three models use specific type of activity nodes (e.g., interaction activities for choreography models). Because the focus of this paper is mainly on GCR decomposition, we abstract their respective formal definitions, but the reader may refer to (21) for more details.

While function $\psi$ maps the activities of the public models to those of the private models, function $\Gamma$ determines the dependencies between the interactions of different public models (e.g., $\Gamma(Request\_details(Special\_carrier)) = Request\_details(Supplier)$). Finally, function $\xi$ represents the dependencies between the activities in the choreography model and those of the public models (e.g., $\xi(order) = \{order(Bulk\_buyer), order(Manufacturer)\}$). Note that in the examples above, connected interaction activities (i.e., the send and the corresponding receive) of different public models have the same labels, while in practice, it is possible to have them different.

Based on functions $\psi$ and $\Gamma$, certain soundness properties of choreography $y$ can be checked, including *structural and behavioral compatibility* between public models, and *consistency* between public and private models (15). Structural consistency requires that for each public activity of the public model of a partner p, there should be a matching element in the corresponding private model of p, but not vice versa (21). Structural compatibility states that for each interaction activity of the public model of a partner p, there should be a matching interaction activity in the public model of another partner. Note that this is a necessary, but not yet sufficient condition for ensuring compatibility and consistency–the models' behaviors (control flow dependencies) should also be compatible and consistent. In this paper, we assume that the choreography $y$ is sound.

In previous work (20), multiple formal languages employed for business process compliance modelling and checking (e.g., linear temporal logic LTL, event calculus EC, extended compliance rule graph eCRG) were compared. It was shown that most of these languages can deal with most qualitative time patterns, and can therefore be used to model the compliance constructs addressed in this paper. Similar results were proven in (30).

In order to specify these constructs as well as transitivity properties required for the GCR decomposition, this work utilizes the visual *eCRG (extended Compliance Rule Graph)* language (42; 38; 36). The eCRG language offers a visual rule notation for expressing compliance rules over process choreographies and is based on first order predicate logic (cf. Fig. 6). To distinguish between a precondition (i.e., antecedence) and corresponding postconditions (i.e., consequences), an eCRG can be partitioned into an *antecedence pattern* and a *consequence pattern*. The antecedence pattern specifies when the compliance rule is triggered (i.e., activated), whereas the consequence pattern specifies what the rule demands. As compliance rules may require the occurrence or absence of certain activities or interactions (i.e., message exchanges), the antecedence and consequence patterns are further sub-divided into occurrence and absence nodes. Sequence conditions between these events can be expressed using directed connectors between the respective nodes. We use the following notation: $\boxed{A}$: Antecedence occurrence; $\boxed{\cancel{A}}$: Antecedence absence; $\text{\textcircled{A}}$: Consequence occurrence; $\text{\textcircled{\cancel{A}}}$: Consequence absence. Fig. 6 introduces the elements of the eCRG language. For a formal definition of eCRG, we refer to Def. 2.

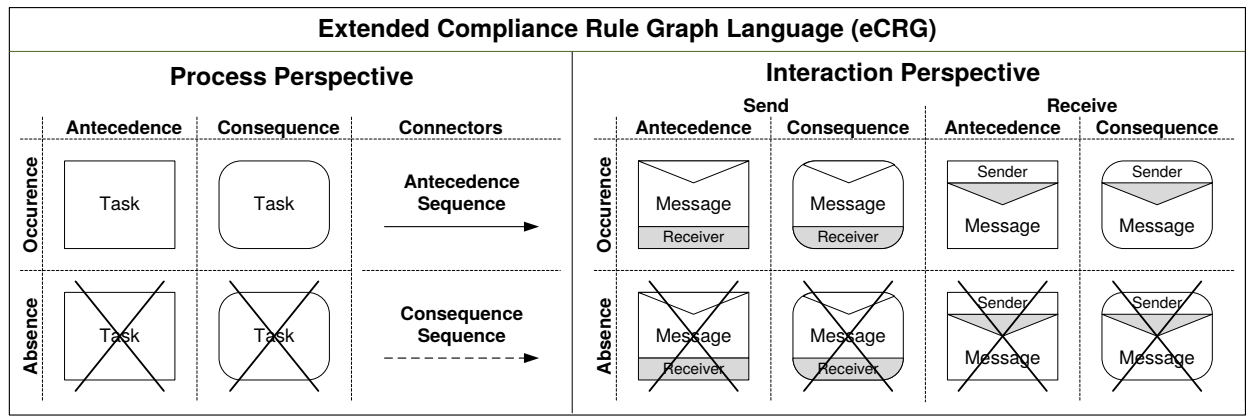

**Figure 6:** Elements of the eCRG language

**Definition 2 (Global Compliance Rule (GCR) structure).** *Given a process choreography y = ($\mathcal{P}, \mathcal{G}, \Pi, \mathcal{L}, \psi, \Gamma, \xi$) (cf. Def. 1), let $\mathcal{A}$ be the set of private and public non-interaction activities and $\mathcal{I}$ be the set of interaction activities. Then: A GCR r is defined as tuple r = ($N, \rho, \varphi, type, pattern$) with*

- *N being the set of* nodes,

- $\rho : N \rightarrow \mathcal{P}$ *returning the partner responsible for a node.*

- $\varphi : N \times N \rightarrow \{ \dashrightarrow, \rightarrow, \emptyset \}$ *returning the sequence flow connector between two nodes, i.e., consequence sequence and antecedence sequence connectors respectively.*

- $type : N \rightarrow \mathcal{A} \cup \mathcal{I}$ *mapping each node to an* activity *or an* interaction *(i.e., message exchange).*

- $pattern : N \rightarrow \{ \boxed{A}, \boxed{\not{A}}, \boxed{C}, \boxtimes \}$

Think of an eCRG as a graph of connected nodes, where each node is assigned to a particular partner (e.g., in C1, $\rho(production) = manufacturer$). A node may either be a private, non-interaction public activity, or an interaction (see Figure 6). Given two nodes of an eCRG, function $\varphi$ returns the sequence flow connector as depicted in Figure 6, where a dashed arrow (i.e., consequence connector) connects an antecedence pattern to set of consequence patterns (e.g, C1: After production a final test is required), and an antecedence connector expresses a relation between antecedence patterns solely (i.e, the pre-condition). For example, assume that we change C3 as follows: $\boxed{Get\_permission\_of\_authority} \rightarrow \boxed{safety\_check} \dashrightarrow \boxed{transport\_intermediate}$. Then: if the pre-condition (i.e., execution of activity $Get\_permission\_of\_authority$ followed by the one of activity $safety\_check$) is met, the post condition (i.e., activity $transport\_intermediate$) will be triggered. Finally, function pattern evaluates whether a node is an antecedence or consequence, and whether or not it should occur.

## 3. Global Compliance Rule Decomposition Theorems

This section introduces the theoretical foundations for the decomposition of global compliance rules (including theorems and proofs) illustrated by a number of examples, which we have extracted from the application scenario introduced in Section 1. Section 3.1 first describes the basic idea of our approach (i.e., why do we need to decompose a GCR), followed by the presentation, proofs and illustrations of the theorems in Section 3.2.

### 3.1. Basic Idea

Our method for the decentralized checking of global compliance rules relies on the decomposition of the original GCR into a set of assertions that can be checked locally by each partner and collectively reproduce the behavior of the GCR (cf. Fig. 7). A communication between partners is only required in the setup phase to deduct the assertions. During runtime, however, no further compliance-related communication becomes necessary for checking the GCR unless a local assertion becomes violated. The decomposition of a GCR into a set of assertions is based on well-grounded theorems, which ensure that if a conjunction of hypotheses is true, the conclusion (GCR) is true as well.

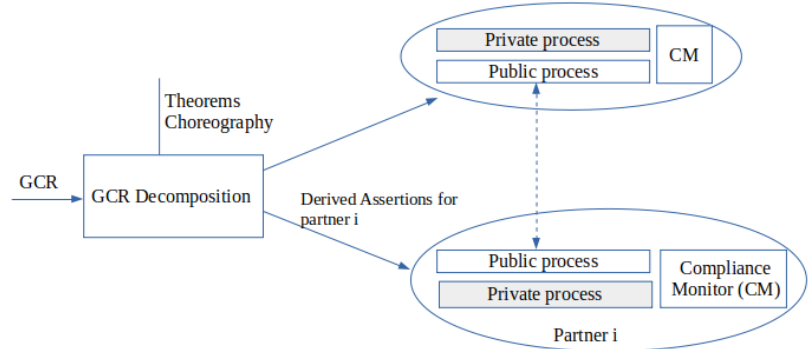

**Figure 7:** Method Overview

## 3.2. Theorems

In this section, we provide a decomposition method for selected global compliance patterns and show how they can be applied in a collaborative setting. In particular, we prove a set of theorems required for ensuring the correctness of our decomposition method. Each theorem represents a possible decomposition of a given compliance pattern.

We illustrate the translation of a GCR into a First Order Logic (FOL) expression using basic equivalences as in Def. 3.

**Definition 3 (Basic Equivalences).** *Based on (41), the following equivalences hold by definition. Predicate $x(t, ty)$ describes that at the point in time t an activity (message) of type ty was executed (i.e., sent or received).*

- $\boxed{A}\dashrightarrow\!\!\text{\textcircled{B}} :\Leftrightarrow \forall a : \big( x(a, A) \rightarrow (\exists b : (x(b, B) \wedge a < b)) \big)$
  $\Leftrightarrow \forall a \exists b : \big(x(a, A) \rightarrow (x(b, B) \wedge a < b)\big)$

- $\text{\textcircled{A}}\dashrightarrow\!\!\boxed{B} :\Leftrightarrow \forall b : \big( x(b, B) \rightarrow (\exists a : (x(a, A) \wedge a < b)) \big)$
  $\Leftrightarrow \forall b \exists a : \big(x(b, B) \rightarrow (x(a, A) \wedge a < b)\big)$

- $\boxed{A}\dashrightarrow\!\!\text{\textcircled{\!\!\!\!\!/\!B}} :\Leftrightarrow \forall a :$
  $\big( x(a, A) \rightarrow (\nexists b : (x(b, B) \wedge a < b)) \big)$
  $\Leftrightarrow \forall a, b : \big(x(a, A) \rightarrow \neg(x(b, B) \wedge a < b)\big)$
  $\Leftrightarrow \forall a, b : \big(x(a, A) \rightarrow (\neg x(b, B) \vee b \leq a)\big)$

- $\text{\textcircled{\!\!\!\!\!/\!A}}\dashrightarrow\!\!\boxed{B} :\Leftrightarrow \forall b :$
  $\big( x(b, B) \rightarrow (\nexists a : (x(a, A) \wedge a < b)) \big)$
  $\Leftrightarrow \forall b, a : \big(x(b, B) \rightarrow \neg(x(a, A) \wedge \neg a < b)\big)$
  $\Leftrightarrow \forall a, b : \big(x(a, A) \rightarrow (\neg x(b, B) \vee b \leq a)\big)$

For example, GCR $\boxed{\text{Production}}\dashrightarrow\!\!\text{\textcircled{Final test}}$ is translated into: $\forall a : \big(x(a, \text{Production}) \rightarrow \exists b : x(b, \text{Final test}) \wedge a < b\big)$. Thereby, relation < expresses a temporal precedence between points in time $a$ and $b$. The decomposition algorithm presented in Section 4 exploits the transitivities for GRC as in Theorem 1. Specifically, by combining transitive relations, where each relation can be checked locally by a single partner, it becomes possible to reconstruct the original GCR behavior. Theorem 1 ensures that the behavior of the derived assertions reproduces the behavior of the GCR, but not vice versa.

**Theorem 1 (Transitivities).**
*Let A, B, and C be three activity or message types. Then:*

a. *The rightwards transitivity holds:*

  $\boxed{A}\dashrightarrow\!\!\text{\textcircled{B}} \wedge \boxed{B}\dashrightarrow\!\!\text{\textcircled{C}} \Rightarrow \boxed{A}\dashrightarrow\!\!\text{\textcircled{C}}$

b. *The leftwards transitivity holds:*

  $\text{\textcircled{A}}\dashrightarrow\!\!\boxed{B} \wedge \text{\textcircled{B}}\dashrightarrow\!\!\boxed{C} \Rightarrow \text{\textcircled{A}}\dashrightarrow\!\!\boxed{C}$

In the following, the correctness of Theorem 1 is proven applying Def. 3.

**Proof 1 (Rightwards Transitivity).**
*Let A, B, and C be three activities or interactions. Then* $\boxed{A}\dashrightarrow\!\!\text{\textcircled{B}} \wedge \boxed{B}\dashrightarrow\!\!\text{\textcircled{C}}$
$:\Leftrightarrow \forall a \exists b : \big(x(a, A) \rightarrow (x(b, B) \wedge a < b)\big) \wedge \forall b \exists c : \big(x(b, B) \rightarrow (x(c, C) \wedge b < c)\big)$
$\Leftrightarrow \forall a \Big(\exists b : \big(x(a, A) \rightarrow (x(b, B) \wedge a < b)\big) \wedge \forall b : \exists c : \big(x(b, B) \rightarrow (x(c, C) \wedge b < c)\big)\Big)$
$\Rightarrow \forall a \exists b, c : \Big(\big(x(a, A) \rightarrow (x(b, B) \wedge a < b)\big) \wedge \big(x(b, B) \rightarrow (x(c, C) \wedge b < c)\big)\Big)$
$\Rightarrow \forall a \Big(\exists b, c : \big((x(a, A) \rightarrow x(b, B) \rightarrow (x(c, C) \wedge a < b < c)\big)\Big)$
$\Rightarrow \forall a \exists c : \Big(x(a, A) \rightarrow (x(c, C) \wedge a < c)\Big) \Rightarrow \boxed{A}\dashrightarrow\!\!\text{\textcircled{C}}$  q.e.d.

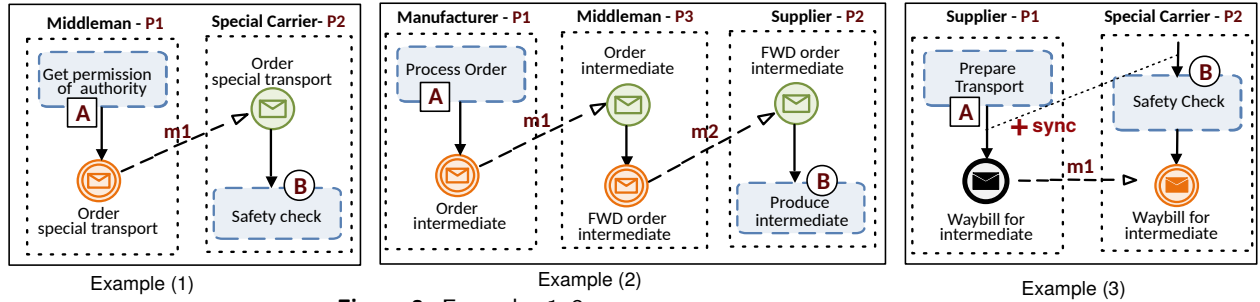

**Figure 8:** Examples 1–3

Leftwards transitivity can be proven similarly by replacing '<' with '>'.

**Corollary 1.** *Let $A$, $B$, $C$, and $D$ be activities or interactions. Then*
$\boxed{A} \dashrightarrow \boxed{B} \land \boxed{B} \dashrightarrow \boxed{C} \land \boxed{C} \dashrightarrow \boxed{D} \Rightarrow \boxed{A} \dashrightarrow \boxed{C} \land \boxed{C} \dashrightarrow \boxed{D} \Rightarrow \boxed{A} \dashrightarrow \boxed{D}$

In the following, we use Examples (1) - (3) (cf. Fig. 8), which we extracted from our running example (cf. Figs. 3 and 4), in order to illustrate how we use Theorem 1 for decomposing a simple compliance rule of type $\boxed{A} \dashrightarrow \boxed{B}$ that involves two private tasks $A$ and $B$ of two different partners $p_1$ and $p_2$ respectively.

- Example (1): $\boxed{get\_permission\_of\_authority} \dashrightarrow \boxed{safety\_check}$. In this example, both activities are private, which would normally require *Middleman* and *Special_carrier* to share runtime information about the execution time of the respective activities. In turn, this would require an agreement on a time synchronization protocol that considers network failures and message transmission delays. This can be solved by identifying a transitive relation between the two private activities that include an interaction activity. According to Theorem 1, the interaction activity *order_special_transport* between *Middleman* and *Special_carrier* fulfills the conditions $A_1$ and $A_2$:
$\boxed{get\_permission\_of\_authority} \dashrightarrow \boxed{order\_special\_transport}$ and $\boxed{order\_special\_transport} \dashrightarrow \boxed{safety\_check}$.
The behavioral and structural compatibility (cf. Section 2) between the partner processes ensures that message *order_special_transport* sent by *Middleman* shall be correctly received by *Special_carrier*. By locally checking $A_1$ and $A_2$ by *Middleman* and *Special_carrier* respectively, we can ensure that the original GCR is not violated as long as the assertions are not violated. If one assertion is violated, a communication between the two partners will become necessary. Note that this violation does not necessarily mean that the original GCR is violated. For example, assume that for a given process instance, assertion $A_1$ evaluates to true, and *Special_carrier* executeś activity *safety_check* before the message arrival. Although this would result in $A_2$ being evaluated to false, it does not necessarily mean that *safety_check* is executed before *get_permission_of_authority*.

- Example (2): $\boxed{process\_order} \dashrightarrow \boxed{produce\_intermediate}$. In this example, *Manufacturer* and *Supplier* do not engage in any direct interaction. However, by looking at the public processes of the collaboration model from Fig. 3, it becomes possible to identify a double transitive relation through *Middleman*, which interacts with both partners. Therefore, using Corollary 1, the transitive relations (assertions): $\boxed{process\_order} \dashrightarrow \boxed{order\_intermediate}$, $\boxed{order\_intermediate} \dashrightarrow \boxed{fwd\_order\_intermediate}$, and $\boxed{fwd\_order\_intermediate} \dashrightarrow \boxed{produce\_intermediate}$ reproduce the behavior of $\boxed{process\_order} \dashrightarrow \boxed{produce\_intermediate}$. *Middleman*, which has initially not been involved in the GCR, becomes involved in the derived assertions. We call *Middleman* an *intermediary partner*.

- Example (3): $\boxed{prepare\_transport} \dashrightarrow \boxed{safety\_check}$. In this example, it is not possible to identify any transitive relation between *Supplier* and *Special_carrier* that involve private activities *prepare_transport* and *safety_check*. The interaction activity *waybill_for_intermediate* connects both partners immediately after the activities in question, which discards any possibility of a transitive relation. In this case, it is not possible to satisfy Theorem 1 and, hence, additional message exchanges become necessary to inform about the execution state of the activities involved in the GCR. Message exchanges can be either synchronous or asynchronous. Asynchronous message exchange only allows for reactive GCR checking and, therefore, detecting violations after their occurrence. Synchronous message exchange, in turn, is proactive as it enforces the GCR with new restrictions to the process models, e.g., the execution of activity *safety_check* becomes enabled only after receiving a synchronization message (i.e., about whether or not *prepare_transport* is executed). *Supplier* shall also inform *Special_carrier* in case activity *prepare_transport* is not executed, as this does not prevent activity *safety_check* from being executed according to the GCR.

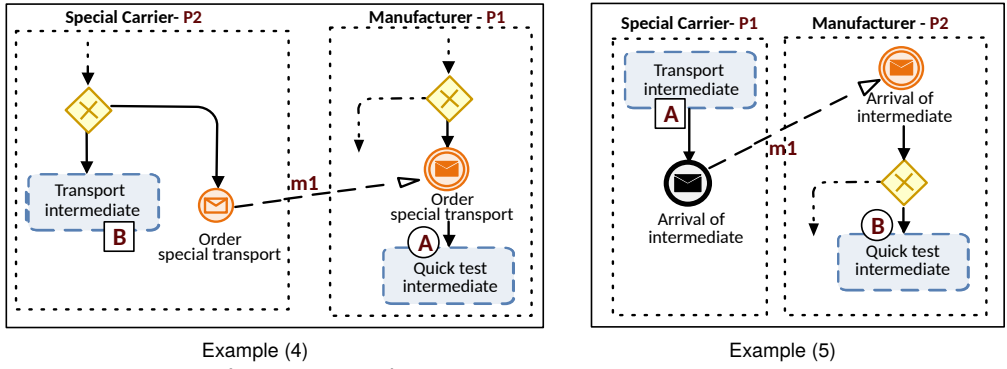

**Figure 9:** Examples 4–5

Rightwards transitivity (cf. Theorem 1.a) directly ensures the correctness of the assertions derived in the above examples. It should be clear that the correctness of the derived assertions in Example (2) can be directly concluded based on Corollary 1. The same examples can be also used to illustrate leftwards transitivity.

**Theorem 2 (Zig zag Transitivities).**
*Let A, B, and C be three activity or message types. Then:*

  a. *The rightwards zig zag transitivity holds for the consequence absence:*
  
  $(B) \dashrightarrow [A] \land [B] \dashrightarrow \cancel{\otimes} \Rightarrow [A] \dashrightarrow \cancel{\otimes}$

  b. *The leftwards zig zag transitivity holds for the consequence absence:*
  
  $[A] \dashrightarrow (B) \land \cancel{\otimes} \dashrightarrow [B] \Rightarrow \cancel{\otimes} \dashrightarrow [A]$

**Proof 2 (Rightwards Zig Zag Transitivity of Absence).**
*Let A, B, and C be activities or interactions. Then:*
$(B) \dashrightarrow [A] \land [B] \dashrightarrow \cancel{\otimes}$

$:\Leftrightarrow \forall a \exists b : \Big( x(a, A) \to (x(b, B) \land b < a) \Big) \land \forall b, c : \Big( x(b, B) \to (\neg x(c, C) \lor c \leq b) \Big)$

$\Leftrightarrow \forall a \Big( \exists b : (x(a, A) \to (x(b, B) \land b < a)) \land \forall b, c : (x(b, B) \to (\neg x(c, C) \lor c \leq b)) \Big)$

$\Rightarrow \forall a \exists b \forall c : \Big( (x(a, A) \to (x(b, B) \land b < a)) \land (x(b, B) \to (x(c, C) \to c \leq b)) \Big)$

$\Rightarrow \forall a \exists b \forall c : \Big( (x(a, A) \to x(b, B) \to (x(c, C) \to c \leq b < a)) \Big)$

$\Rightarrow \forall a \forall c : \Big( x(a, A) \to (x(c, C) \to c \leq a) \Big)$

$\Rightarrow \forall a, c : \Big( x(a, A) \to (\neg x(c, C) \lor c \leq a) \Big) \Rightarrow [A] \dashrightarrow \cancel{\otimes}$   q.e.d.

Leftwards *zig zag* transitivity of absence can be proven similarly by replacing '<' with '>' and '≤' with '≥'.

In the following, we use Examples (4) and (5) from Fig. 9 to illustrate and discuss how Theorem 2 can be used to decompose a GCR of type rightwards zigzag $[A] \dashrightarrow \cancel{(B)}$. Note that these two examples are adopted from the running example we introduced in Section 1 in order to fulfill the decomposition requirements.

- Example (4): $[quick\_test\_intermediate] \dashrightarrow \cancel{(transport\_intermediate)}$. In this example, *transport_intermediate* and *order_special_transport* in *Special_carrier* belong to different XOR branches, which means that the execution of activity *transport_intermediate* implies the non-execution of activity *order_special_transport* and vice versa (fulfilling assertion $A_1$ ⟨order_special_transport⟩ ⇢ ⟨transport_intermediate⟩). Additionally, in *Manufacturer*, the interaction activity *order_special_transport* and the private activity *quick_test_intermediate* belong to the same XOR branch, and fulfill assertion $A_2$ ⟨order_specialtransport⟩ ⇢ ⟨quick_test_intermediate⟩. According to Theorem 2.a, the conjunction of $A_1$ and $A_2$ reproduces the behavior of the original GCR. Note that process compatibility ensures that whenever sending *order_special_transport* occurs in *Special_carrier*, receiving *order_special_transport* should occur in *Manufacturer* as well. At the presence of loops that encapsulate the depicted process part of *Special_carrier*, the XOR fragment can be executed multiple times possibly leading to an alternate execution of the corresponding branches. For example, if in the first loop iteration, *transport_intermediate* is executed and *quick_test_intermediate* is not executed, then, to this point both derived assertions are satisfied. Let us assume that a future iteration over the XOR fragment in the context of the same process instance triggers *quick_test_intermediate* execution, thus, violating the original GCR.

iteration 1: *special_carrier*: {*transport_intermediate*}
iteration 1: *manufacturer* { }
iteration 2: *special_carrier* {*transport_intermediate*, **order_special_transport**}
iteration 2: *manufacturer* {*order_special_transport*, **quick_test_intermediate**}
Combined trace: {**transport_intermediate**, *order_special_transport*, *order_special_transport*, *order_special_transport*, **quick_test_intermediate**}

By looking at the combined trace, it becomes clear that the GCR is violated. Unfortunately, this would require both partners to exchange the traces and use a common time stamping system to obtain the same chronological order of activities. Using the theorems, however, the *Manufacturer* can locally run its derived assertion against its own execution trace of the same process instance, and identify the violation. Note that the decomposition does not enforce the processes with new restrictions (except when no transitivity could be derived), but uses the existing control flow and interactions between partners to derive assertions that can be used for a decentralized checking of the original GCR.

- Example (5): ⟨quick_test_intermediate⟩ ⇢ *transport_intermediate*. In Fig. 9, *quick_test_intermediate* always happens after *arrival_of_intermediate* ensuring ⟨quick_test_intermediate⟩ ⇢ *arrival_of_intermediate*. The second part of the decomposition can be directly derived from the process control flow of *Special_carrier*: *transport_intermediate* ⇢ *arrival_of_intermediate*. The same reasoning applies to this example at the presence of loops. The correctness of Example (5) concludes from the leftwards zig zag transitivity (cf. Theorem 2.b), whereas Example (4) relies on the rightwards zig zag transitivity of the absence (cf. Theorem 2.a). The decomposition process is not limited to these scenarios and, as aforementioned, the decomposition cannot always be automated, but might require manual interaction and processing. Altogether, the decomposition eases global compliance rule checking, where each partner checks its corresponding derived assertions locally. A GCR is rechecked only if at least one of the derived assertions is not evaluated to true. Note that this does not necessarily imply that the GCR is violated.

**Theorem 3 (Rightwards Chaining Transitivity).**
Let $A, B, C,$ and $D$ be activities or interactions such as $A \to B \dashrightarrow C \dashrightarrow D$: if A and B occur, C and D shall occur afterwards. Let $m_1, m_2,$ and $m_3$ be three interactions such as:

(1) $M_1 \dashrightarrow A$

(2) $M_1 \to B \dashrightarrow M_2$

(3) $M_2 \dashrightarrow C \dashrightarrow M_3$

(4) $M_3 \dashrightarrow D$

Then: Whenever $(1) \land (2) \land (3) \land (4)$ evaluates to true, $A \to B \dashrightarrow C \dashrightarrow D$ is true as well.

**Proof 3 (Rightwards Chaining Transitivity).**
$(1) \land (2)$ :
$:\Leftrightarrow \quad \forall a \exists m_1 \ : \ \Big(x(a, A) \to x(m_1, M_1) \land (m_1 < a)\Big) \land \forall m_1, \forall b, \exists m_2 \ : \ \Big((x(m_1, M_1) \land x(b, B) \land m_1 < b) \to$

$(x(m_2, M_2) \wedge b < m_2)\Big)$

$\vDash \forall a \exists m_1 \forall b \exists m_2 : \Big( x(a, A) \rightarrow x(m_1, M_1) \wedge (m_1 < a) \Big) \wedge \Big( (x(m_1, M_1) \wedge x(b, B) \wedge m_1 < b) \rightarrow (x(m_2, M_2) \wedge b < m_2) \Big)$

$\vDash \forall a \exists m_1 \forall b \exists m_2 : \Big( x(a, A) \wedge x(b, B) \wedge a < b \rightarrow x(m_1, M_1) \wedge (m_1 < b) \Big) \rightarrow (x(m_2, M_2) \wedge b < m_2)$

$\vDash \forall a \forall b \exists m_2 : x(a, A) \wedge x(b, B) \wedge a < b \rightarrow x(m_2, M_2) \wedge b < m_2$

$(1) \wedge (2) \wedge (3)$

$\vDash \forall a \forall b \exists m_2 : \Big( x(a, A) \wedge x(b, B) \wedge a < b \rightarrow x(m_2, M_2) \wedge b < m_2 \Big) \wedge \forall m_2 \exists c \exists m_3 : \Big( x(m_2, M_2) \rightarrow x(c, C) \wedge x(m_3, M_3) \wedge m_2 < c \wedge c < m_3 \Big)$

$\vDash \forall a \forall b \exists c \exists m_3 : x(a, A) \wedge x(b, B) \wedge a < b \rightarrow x(c, C) \wedge x(m_3, M_3) \wedge b < c \wedge c < m_3$

$(1) \wedge (2) \wedge (3) \wedge (4)$

$\vDash \forall a \forall b \exists c \exists m_3 : \Big( x(a, A) \wedge x(b, B) \wedge a < b \rightarrow x(c, C) \wedge x(m_3, M_3) \wedge b < c \wedge c < m_3 \Big) \wedge \forall m_3 \exists d : \Big( x(m_3, M_3) \rightarrow x(d, D) \wedge m_3 < d \Big)$

$\vDash \forall a \forall b \exists c \exists d : \Big( x(a, A) \wedge x(b, B) \wedge a < b \rightarrow x(c, C) \wedge x(d, D) \wedge b < c \wedge c < d \Big)$

- Example (6): $\boxed{get\_permission\_authority} \rightarrow \boxed{prepare\_transport} \dashrightarrow \boxed{transport\_intermediate} \dashrightarrow \boxed{production}$. In this example, all activities involved in the GCR are private and belong to separate partners. According to the process models shown in Figs. 3 and 4, each partner can separately derive the corresponding assertion based on Theorem 3 and involving the corresponding activity in the GCR.

    (1) $Middleman$ $\boxed{fwd\_order\_intermediate} \dashrightarrow \boxed{get\_permission\_authority}$

    (2) $Supplier$ $\boxed{fwd\_order\_intermediate} \rightarrow \boxed{prepare\_transport} \dashrightarrow \boxed{waybill\_for\_intermediate}$

    (3) $Special\_carrier$ $\boxed{waybill\_for\_intermediate} \dashrightarrow \boxed{transport\_intermediate} \dashrightarrow \boxed{arrival\_of\_intermediate}$

    (4) $Manufacturer$ $\boxed{arrival\_of\_intermediate} \dashrightarrow \boxed{production}$

    In this example, partners will first engage in a setup phase, in which they agree on the interaction activities that satisfy all derived assertions following the assertions' templates of Theorem 3. For example, $Middleman$ will start by identifying relations in its process of type $\boxed{interaction\_activity} \dashrightarrow \boxed{get\_permission\_authority}$, where $interaction\_activity$ must be a message exchange with $Supplier$ that is the partner being responsible for the following antecedence occurrence $prepare\_transport$. In this example, $Middleman$ and $Supplier$ have only one interaction that satisfies the derived assertion (1); however, it is also possible to identify several alternatives. The combination of the four derived assertions reproduce the behavior of the original GCR when all assertions are true.

The following theorem represents a generalization of Theorem 3 with $n$ antecedences' occurrences and $m$ consequences' occurrences. Note that the previous example also illustrates Theorem 4 with $n = 2$ and $m = 2$.

**Theorem 4 (Generic Rightwards Chaining Transitivity).**
Let $A_{1 \leq i \leq n}$ and $C_{1 \leq j \leq m}$ be $n+m$ activities. $\boxed{A_1} \rightarrow ... \rightarrow \boxed{A_n} \dashrightarrow \boxed{C_1} \rightarrow ... \rightarrow \boxed{C_m}$: if all $A_i$ occur such that $\forall i < n, A_i < A_{i+1}$ holds, then all $C_j$ should occur afterwards such that $\forall j < m, C_j < C_{j+1}$ holds.
Let $m_k$, where $1 < k < n+m-1$ be interactions such that:

(1) $\boxed{M_1} \dashrightarrow \boxed{A_1}$

(2) $\boxed{M_{i-1}} \rightarrow \boxed{A_i} \dashleftarrow \boxed{M_i}$ where $1 < i < n$

(3) $\boxed{M_{n-1}} \rightarrow \boxed{A_n} \dashrightarrow \boxed{M_n}$

(4) $\boxed{M_{n+j-1}} \dashrightarrow \boxed{C_j} \dashrightarrow \boxed{M_{n+j}}$ where $1 \leq j < m$

(5) $\boxed{M_{n+m-1}} \dashrightarrow \boxed{C_m}$

Then: Whenever (1) ∧ (2) ∧ (3) ∧ (4) ∧ (5) *evaluates to true*, $\boxed{A_1} \dashrightarrow ... \to \boxed{A_n} \dashrightarrow \boxed{C_1} \dashrightarrow ... \dashrightarrow \boxed{C_m}$ *is true as well*.

**Proof 4 (Generic Rightwards Chaining Transitivity).**
(1)
$:\Leftrightarrow \forall a \exists m_1 : \left( x(a, A) \to x(m_1, M_1) \land (m_1 < a) \right)$
(2)
$:\Leftrightarrow \bigwedge_{1 < i < n} \left( \forall m_{i-1} \forall a_i \exists m_i : (x(m_{i-1}, M_{i-1}) \land x(a_i, A_i) \land m_{i-1} < a_i) \to (x(m_i, M_i) \land m_i < a_i) \right)$
(3)
$:\Leftrightarrow \left( \forall m_{n-1} \forall a_n \exists m_n : (x(m_{n-1}, M_{n-1}) \land x(a_n, A_n) \land m_{n-1} < a_n) \to (x(m_n, M_n) \land a_n < m_n) \right)$

(1) ∧ (2) ∧ (3): *when evaluated to true, this ensures that all $A_i$ were executed, and all these executions combined lead to $M_n$ as a consequence. This includes the case where all $A_i$ execute in ascending order. So, if we consider this particular order, the formula leading to $M_n$ becomes true.*

(4)
$:\Leftrightarrow \bigwedge_{1 \leq j < m} \left( \forall m_{n+j-1} \forall c_j \exists m_{n+j} : (x(m_{n+j-1}, M_{n+j-1}) \land x(c_j, C_j) \land m_{n+j-1} < c_n) \to (x(m_{n+j}, M_{n+j}) \land m_{n+j} < c_j) \right)$
(5)
$:\Leftrightarrow \left( \forall m_{n+m-1} \exists c_m : (x(m_{n+m-1}, M_{n+m-1}) \to (x(c_m, C_m) \land m_{n+m-1} < c_m) \right)$

(4)∧(5): *This formula transitively ensures that whenever $M_{n+j-1}$ is executed, there is a least one execution in ascending order of all $C_j$. $M_{n+j-1}$ becomes the link between all antecedence patterns and consequence patterns. Therefore, the conjunction of formulas (1) to (5) ensures that whenever an instance containing an ordered execution of $A_i$ should lead to an ordered execution of execution of $C_i$. Note that this conjunction represents a stronger constraint than the original GCR. However, as these formulas are deducted directly from the actual processes, they do not add new constraints.*

**Theorem 5 (Between Pattern 1).**
Let A, B and C be three activities. $\boxed{A} \dashrightarrow \boxed{C} \dashrightarrow \boxed{B}$: *if A and B occur and B occurs after A, then C must occur in between*.

(1) $\boxed{A} \dashrightarrow \boxed{M_1} \oplus \cancel{\boxed{M_2}} \dashrightarrow \boxed{M_1}$

(2) $\boxed{M_2} \dashrightarrow \boxed{M_3} \dashrightarrow \boxed{B}$

(3) $\boxed{M_1} \dashrightarrow \boxed{C} \dashrightarrow \boxed{M_3}$

Then: Whenever (1) ∧ (2) ∧ (3) *evaluates to true*, $\boxed{A} \dashrightarrow \boxed{C} \dashrightarrow \boxed{B}$ *is true as well*.

**Proof 5 (Between Pattern 1).**
(1) $:\Leftrightarrow \forall a, x(a, A) \to \exists m_1 \nexists m_2, x(m_1, M_1) \land x(m_2, M_2) \land (a < m_1) \land (m_2 < m_1)$
(2) $:\Leftrightarrow \forall b, x(b, B) \to \exists m_2 \exists m_3, x(m_2, M_2) \land x(m_3, M_3) \land (m_2 < m_3 < b)$
(3) $:\Leftrightarrow \forall m_1 \forall m_3, x(m_1, M_1) \land x(m_3, M_3) \to \exists c, x(c, C) \land (m_1 < c < m_3)$
(1)∧(2) $\Leftrightarrow \forall a \forall b, x(a, A), x(b, B) \to (\exists m_1 \nexists m_2, x(m_1, M_1) \land x(m_2, M_2) \land (a < m_1) \land (m_2 < m_1)) \land (\exists m_2 \exists m_3, x(m_2, M_2) \land x(m_3, M_3) \land (m_2 < m_3 < b))$
$\vDash \forall a \forall b, x(a, A) \land x(b, B) \to \exists m_1 \exists m_2 \exists m_3, x(m_1, M_1) \land x(m_2, M_2) \land x(m_3, M_3) \land (a < m_1 < m_2) \land (m_2 < m_3 < b))$
$\vDash \forall a \forall b, x(a, A) \land x(b, B) \land a < b \to \exists m_1 \exists m_2 \exists m_3, x(m_1, M_1) \land x(m_2, M_2) \land x(m_3, M_3) \land (a < m_1 < m_2 < m_3 < b))$
(1) ∧ (2) ∧ (3) $:\vDash \forall a \forall b, x(a, A) \land x(b, B) \land a < b \to \exists m_1 \exists m_2 \exists m_3, x(m_1, M_1) \land x(m_2, M_2) \land x(m_3, M_3) \land (a < m_1 < m_2 < m_3 < b)) \to \exists c, x(c, C) \land (m_1 < c < m_3)$
$\vDash \forall a \forall b \exists m_1 \exists m_2 \exists m_3 \exists c, x(a, A) \land x(b, B) \land a < b \to x(m_1, M_1) \land x(m_2, M_2) \land x(m_3, M_3) \land x(c, C) \land (a < m_1 < m_2 < m_3 < b)) \land (m_1 < c < m_3)$
$\vDash \forall a \forall b, x(a, A) \land x(b, B) \land a < b \to \exists c, x(c, C) \land (a < c < b))$

- Example (7): $\boxed{order\_intermediate} \dashrightarrow \boxed{prepare\_transport} \dashrightarrow \boxed{transport\_intermediate}$. Again, in this example, we consider the worst case scenario where each activity referred to by the GCR belongs to a different process partner. In this example, *Middleman* has one single alternative as it only has two interaction activities with *Special_carrier* and *Supplier* respectively, which follow the assertion template (1) of Theorem 5; i.e., each execution of *order_intermediate* must be followed (not necessarily immediately) by *fwd_order_intermediate*, which, in turn, should not be preceded by any *order_special_transport* execution. Similarly, *Special_carrier* and *Supplier* should identify assertions that follow rule templates (2) and (3) respectively.

Note that the Between Pattern $\boxed{A} \dashrightarrow \boxed{C} \dashrightarrow \boxed{B}$ can be also checked using chaining transitivity $\boxed{A} \dashrightarrow \boxed{C} \dashrightarrow \boxed{B}$. However, this adds a stronger assumption on $C$ and $B$ that should follow $A$ whenever it occurs. For example, this holds in the running example (cf. Fig. 3), as *order_intermediate* transitively implies *prepare_transport*, which, in turn, transitively implies *transport_intermediate*.

(1) $Middleman$ $\boxed{order\_intermediate} \dashrightarrow \boxed{fwd\_order\_intermediate}$
$\oplus$ $\boxed{order\_special\_transport} \dashrightarrow \boxed{fwd\_order\_intermediate}$

(2) $Special\_carrier$ $\boxed{order\_special\_transport} \dashrightarrow \boxed{waybill\_for\_intermediate} \dashrightarrow \boxed{transport\_intermediate}$

(3) $Supplier$ $\boxed{fwd\_order\_intermediate} \dashrightarrow \boxed{prepare\_transport} \dashrightarrow \boxed{waybill\_for\_intermediate}$

**Theorem 6 (Between Pattern 2).**
Let A, B and C be three activities. $\boxed{A} \dashrightarrow \boxed{C} \dashrightarrow \boxed{B}$: if A and B occur and B occurs after A, then C shall occur in between.

*(1)* : $\boxed{M_1} \dashrightarrow \boxed{A} \dashrightarrow \boxed{M_2} \dashrightarrow \boxed{M_3} \dashrightarrow \boxed{M_4}$

*(2)* : $\boxed{M_1} \dashrightarrow \boxed{B} \dashrightarrow \cancel{\boxed{M_3}} \dashrightarrow \boxed{M_4}$

*(3)* : $\boxed{M_3} \dashrightarrow \boxed{M_5} \dashrightarrow \boxed{B}$

*(4)* : $\boxed{M_2} \dashrightarrow \boxed{C} \dashrightarrow \boxed{M_5}$

Then: Whenever $(1) \land (2) \land (3) \land (4)$ evaluates to true, $\boxed{A} \dashrightarrow \boxed{C} \dashrightarrow \boxed{B}$ is true as well.

**Proof 6 (Between Pattern 2).**
$(1) :\Leftrightarrow \forall a, x(a, A) \rightarrow \exists m_1 \exists m_2 \exists m_3 \exists m_4, x(m_1, M_1) \land x(m_2, M_2) \land x(m_3, M_3) \land x(m_4, M_4) \land (m_1 < a < m_2 < m_3 < m_4)$
$(2) :\Leftrightarrow \forall m_1 \forall b \forall m_4, x(m_1, M_1) \land x(b, B) \land x(m_4, M_4) \rightarrow \nexists m_2, x(m_2, M_2) \land (b < m_2 < m_4)$
$(3) :\Leftrightarrow \forall m_3 \forall b, x(m_3, M_3) \land x(b, B) \rightarrow \exists m_5, x(m_5, M_5) \land (m_3 < m_5 < b)$
$(4) :\Leftrightarrow \forall m_2 \forall m_5, x(m_2, M_2) \land x(m_5, M_5) \rightarrow \exists c, x(c, c) \land (m_2 < c < m_5)$

*Using (1):*
$\vDash \forall a \forall b, x(a, A) \land x(b, B) \land a < b \rightarrow \exists m_1 \exists m_2 \exists m_3 \exists m_4, x(m_1, M_1) \land x(m_2, M_2) \land x(m_3, M_3) \land x(m_4, M_4) \land (m_1 < a < m_2 < m_3 < m_4) \land (a < b < m_4 \lor m_4 < b)$

$\vDash \forall a \forall b \exists m_1 \exists m_2 \exists m_3 \exists m_4, x(a, A) \land x(b, B) \land a < b \rightarrow \Big(x(m_1, M_1) \land x(m_2, M_2) \land x(m_3, M_3) \land x(m_4, M_4)\Big) \land \Big(((m_1 < a < m_2 < m_3 < m_4) \land (m_1 < b < m_4)) \lor ((m_1 < a < m_2 < m_3 < m_4) \land (m_4 < b))\Big)$

*Using (2), if b happens before $m_4$ then $m_3$ should not happen in between:*
$\vDash \forall a \forall b \exists m_1 \exists m_2 \exists m_3 \exists m_4, x(a, A) \land x(b, B) \land a < b \rightarrow \Big(x(m_1, M_1) \land x(m_2, M_2) \land x(m_3, M_3) \land x(m_4, M_4)\Big) \land \Big((m_1 < a < m_2 < m_3 < b < m_4) \lor (m_1 < a < m_2 < m_3 < m_4 < b)\Big)$

$\vDash \forall a \forall b \exists m_2 \exists m_3, x(a, A) \land x(b, B) \land a < b \rightarrow \Big(x(m_2, M_2) \land x(m_3, M_3) \land (a < m_2 < m_3 < b)\Big)$

*Using (3), if $m_3 < b$, then there should be $m_4$ in between:*
$\vDash \forall a \forall b \exists m_2 \exists m_3, x(a, A) \land x(b, B) \land a < b \rightarrow \Big(x(m_2, M_2) \land x(m_3, M_3) \land (a < m_2 < m_3 < b) \rightarrow \exists m_5, x(m_5, M_5) \land (m_3 < m_5 < b)\Big)$

$\vDash \forall a \forall b \exists m_2 \exists m_3 \exists m_5, x(a, A) \land x(b, B) \land a < b \rightarrow \Big(x(m_2, M_2) \land x(m_3, M_3) \land x(m_5, M_5)) \land (a < m_2 < m_3 < m_5 < b)\Big)$

$\vDash \forall a \forall b \exists m_2 \exists m_3 \exists m_5, x(a, A) \land x(b, B) \land a < b \rightarrow \Big(x(m_2, M_2) \land x(m_5, M_5)) \land (a < m_2 < m_5 < b)\Big)$

*Using (4), if there exist $m_2$ and $m_5$ such that $m_2 < m_5$, then there should be c in between:*
$\vDash \forall a \forall b \exists m_2 \exists m_5, x(a, A) \land x(b, B) \land a < b \rightarrow \Big(x(m_2, M_2) \land x(m_5, M_5)) \land (a < m_2 < m_5 < b) \rightarrow \exists c, x(c, C) \land m_2 < c < m_5\Big)$

⊨ $\forall a \forall b \exists m_2 \exists m_5 \exists c, x(a, A) \land x(b, B) \land a < b \rightarrow \big( x(m_2, M_2) \land x(m_5, M_5) \land x(c, C)) \land (a < m_2 < c < m_5 < b) \big)$

⊨ $\forall a \forall b \exists c, x(a, A) \land x(b, B) \land a < b \rightarrow x(c, C)) \land (a < c < b)$

In order to illustrate Theorem 6, we apply the following adaptations to the running example (cf. Figs. 1 - 4):
**(i)** After receiving *request_details*, *Supplier* prepares the details privately (*prepare_details*), then informs *Middleman* about the start of intermediate production (*production_status*) before sending back *transport_details*.
**(ii)** After receiving transport details, *Special_carrier* confirms to *Middleman* the availability of transportation for intermediate (*transport_confirmation*).
**(iii)** After receiving *order_special_transport*, *Middleman* receives *production_status*, does *internal_checks*, and waits for *transport_confirmation*.

- Example (8): [prepare_details]⇢(internal_checks)⇢[safety_check]. Similar to the previous examples, the partners start by locally identifying relations that satisfy the derived assertions templates of Theorem 6, then apply a matching mechanism to check whether the additional interactions used for the derived assertions intersect and jointly fulfill the templates. It is noteworthy that the number of additional interaction activities required for the derived assertions is superior to the number required in Theorem 5. Despite that, Theorem 6 provides more relaxed assumptions compared to Theorem 5 as it does not restrict activity *B* from occurring before activity *A*. Theorem 6 still prevents *B* from happening between $m_1$ and $m_3$. The following assertions are the decomposition results of Example (8):

  (1) *Supplier* : [request_details]⇢[prepare_details]⇢(production_status)⇢(transport_details)⇢(waybill_intermediate)
  (2) *Special_carrier* : [request_details]⇢[safety_check]⇢~~(transport_details)~~⇢[waybill_intermediate]
  (3) *Special_carrier* : [transport_details]⇢(transport_confirmation)⇢[safety_check]
  (4) *Middleman* : [production_status]⇢(internal_checks)⇢[transport_confirmation]

Note that all previous theorems consider loops and multiple occurrences of each of the activities composing the global compliance rule GCR. Indeed, in a process model that includes loops or multiple instance patterns, an activity may be executed multiple times at different points in time in the context of one single process instance. As such, each derived assertion including such repetitive activity should be satisfied for all its occurrences. Although this issue has been addressed in all previous theorems (see proofs), it resulted in additional decomposition complexity not required for loop-free processes. Therefore, we propose a simpler decomposition method for the "between" pattern, which may be applied solely to loop-free processes.

**Theorem 7 (Between Pattern (without loops)).**
*Let A, B and C be three activities.* [A]⇢(C)⇢[B] : *if A and B occur and B occurs after A, then C shall occur in between.*

- (1) : [A]⇢(M_1)⇢(M_2)
- (2) : (M_2)⇢(M_3)⇢[B]
- (3) : (M_1)⇢(C)⇢(M_3)

*Then: If the conjunction of formulas (1) ∧ (2) ∧ (3) evaluates to true,* [A]⇢(C)⇢[B] *is true as well.*

**Proof 7 (Between Pattern (without loops)).**
(1) :⇔ $\forall a : (x(a, A) \rightarrow \exists m_1, \exists m_2 : (x(m_1, M_1) \land x(m_2, M_2) \land (a < m_1 < m_2))$
⇔ $\forall a \exists m_1, \exists m_2 : (x(a, A) \rightarrow (x(m_1, M_1) \land x(m_2, M_2) \land (a < m_1 < m_2))$
(2) :⇔ $\forall b : (x(b, B) \rightarrow \exists m'_2, \exists m_3 : (x(m'_2, M_2) \land x(m_3, M_3) \land (m'_2 < m_3 < b))$
(2) :⇔ $\forall b \exists m'_2, \exists m_3 : (x(b, B) \rightarrow (x(m'_2, M_2) \land x(m_3, M_3) \land (m'_2 < m_3 < b))$
(1) ∧ (2) ⊨ $(\forall a \forall b \exists m_1, \exists m_2 \exists m'_2 \exists m_3 : (x(a, A) \land (x(b, B)) \rightarrow (x(m_1, M_1) \land x(m_2, M_2) \land x(m'_2, M_2) \land x(m_3, M_3) \land (a <$

$m_1 < m_2) \wedge (m'_2 < m_3 < b))$

As there are no loops over A, B and corresponding $M_i$ messages, then $M_2$ can occur only once within one process instance. Consequently, $m_2 = m'_2$:
(1) $\wedge$ (2) $\vDash$ $(\forall a \forall b \exists m_1, \exists m_2 \exists m_3 : (x(a,A) \wedge (x(b,B)) \rightarrow (x(m_1, M_1) \wedge x(m_2, M_2) \wedge x(m_3, M_3) \wedge (a < m_1 < m_2 < m_3 < b))$
$\vDash (\forall a \forall b \exists m_1, \exists m_3 : (x(a,A) \wedge (x(b,B)) \rightarrow (x(m_1, M_1) \wedge x(m_3, M_3) \wedge (a < m_1 < m_3 < b))$
(3) $:\Leftrightarrow \forall m_1 \forall m_3 : (x(m_1, m_1) \wedge x(m_3, M_3) \rightarrow \exists c : (x(m_1, M_1) \wedge x(c, C) \wedge (m_1 < c < m_3))$
$\Leftrightarrow \forall m_1 \forall m_3 \exists c : (x(m_1, m_1) \wedge x(m_3, M_3) \rightarrow (x(m_1, M_1) \wedge x(c, C) \wedge (m_1 < c < m_3))$
(1) $\wedge$ (2) $\wedge$ (3) $\vDash$ $(\forall a \forall b \exists m_1, \exists m_3 : (x(a,A) \wedge (x(b,B)) \rightarrow (x(m_1, M_1) \wedge x(m_3, M_3) \wedge (a < m_1 < m_3 < b)) \rightarrow \exists c, x(c, C) \wedge (a < m_1 < c < m_3 < b)$
$\vDash (\forall a \forall b(x(a,A) \wedge (x(b,B)) \rightarrow \exists c, x(c, C) \wedge (a < c < b)$

- Example (9): $\boxed{prepare\_details} \dashrightarrow \boxed{internal\_checks} \dashrightarrow \boxed{safety\_check}$. We use the same GCR as in previous theorem illustration (including the adaptations to Fig. ??). As the tasks involved in the GCR are not contained in any loop, Theorem 7 may be applied. The following assertions are the decomposition results:

  (1) $Supplier$ $\boxed{prepare\_details} \dashrightarrow \boxed{production\_status} \dashrightarrow \boxed{transport\_details}$
  
  (2) $Special\_carrier$ $\boxed{transport\_details} \dashrightarrow \boxed{transport\_confirmation} \dashrightarrow \boxed{safety\_check}$
  
  (4) $Middleman$ $\boxed{production\_status} \dashrightarrow \boxed{internal\_checks} \dashrightarrow \boxed{transport\_confirmation}$

**Theorem 8 (Requires transitivity).**
Let A and B be two activities or interactions such as $\boxed{A}$ $\boxed{B}$ : if A occurs then B should occur (before or after, $\forall a, x(a, A) \rightarrow \exists B, x(b, B))$:.
Let A, B, M be three activities or interactions such that :

(1) : $\boxed{A} \dashrightarrow \boxed{M}$

(2) : $\boxed{M}$ $\boxed{B}$.

If (1) $\wedge$ (2) evaluates to true, then $\boxed{A} \dashrightarrow \boxed{B} \vee \boxed{B} \dashrightarrow \boxed{A}$ evaluates to true.

**Proof 8 (Requires transitivity).**
(1) $:\Leftrightarrow \forall a, x(a, A) \rightarrow \exists m : x(m, M) \wedge (a < m)$
(2) $:\Leftrightarrow \forall m, \left(x(m, M) \rightarrow \exists b : x(b, B)) \wedge m < b\right) \vee \forall m, \left(x(m, M) \rightarrow \exists b : x(b, B)) \wedge (b < m)\right)$
$\Leftrightarrow \forall m, \left(x(m, M) \rightarrow \exists b : x(b, B)) \wedge (m < b) \vee b < m)\right)$
(1) $\wedge$ (2) $:\Leftrightarrow \forall a, \left(x(a, A) \rightarrow \exists m : x(m, M) \wedge (a < m)\right) \wedge \forall m, \left(x(m, M) \rightarrow \exists b : x(b, B)) \wedge (m < b) \vee b < m)\right)$
$\vDash \forall a \exists m \exists b, x(a, A) \rightarrow x(m, M) \wedge (a < m) \rightarrow x(b, B) \wedge (m < b) \vee b < m$
$\vDash \forall a \exists m \exists b, x(a, A) \rightarrow x(m, M) \wedge (a < m) \rightarrow x(b, B) \wedge (a < m < b) \vee (a < b < m) \vee (b < a < m)$
$\vDash \forall a \exists b, x(a, A) \rightarrow x(b, B) \wedge (a < b) \vee (b < a)$
$\vDash \forall a, x(a, A) \rightarrow \exists b, x(b, B)$

Note that this theorem also considers loops and multi-instance patterns. The illustration of Theorem 8 is similar to the rightwards and leftwards transitivity examples.

## 4. Algorithm for Decomposing Global Compliance Rules

At design time, checking a GCR that solely refers to interactions and/or public activities can be achieved by applying contemporary compliance checking techniques (cf. (29)) either on the choreography model or the public process models of the involved partners. By contrast, if a GCR refers to private activities of different partners, it becomes impossible to check its correctness as partners must not view the private process model parts of the other partners and, therefore, cannot identify the dependencies between the private activities involved in the GCR. To cope with this issue, we suggest decomposing the GCR into a set of assertions of which each can be checked locally by the corresponding partner. The decomposed rules then reproduce the behavior of the original GCR.

Decomposition in compliance checking has been exploited by (60), but only at the process model level in order to achieve performance gains for the compliance checks. The article at hand proposes to decompose the GCR to distribute the compliance checks to the partners for maintaining the confidentiality of their private tasks. This section focuses on the decomposition algorithms and explains the steps to derive assertions by applying the theorems introduced in Section 3.

Figure 10 provides a high-level description of the steps required by partners involved in a GCR to identify a valid decomposition. Algorithm 1 provides a more detailed view on how this can be achieved in practice with a particular focus on compliance rules that include one antecedence pattern.

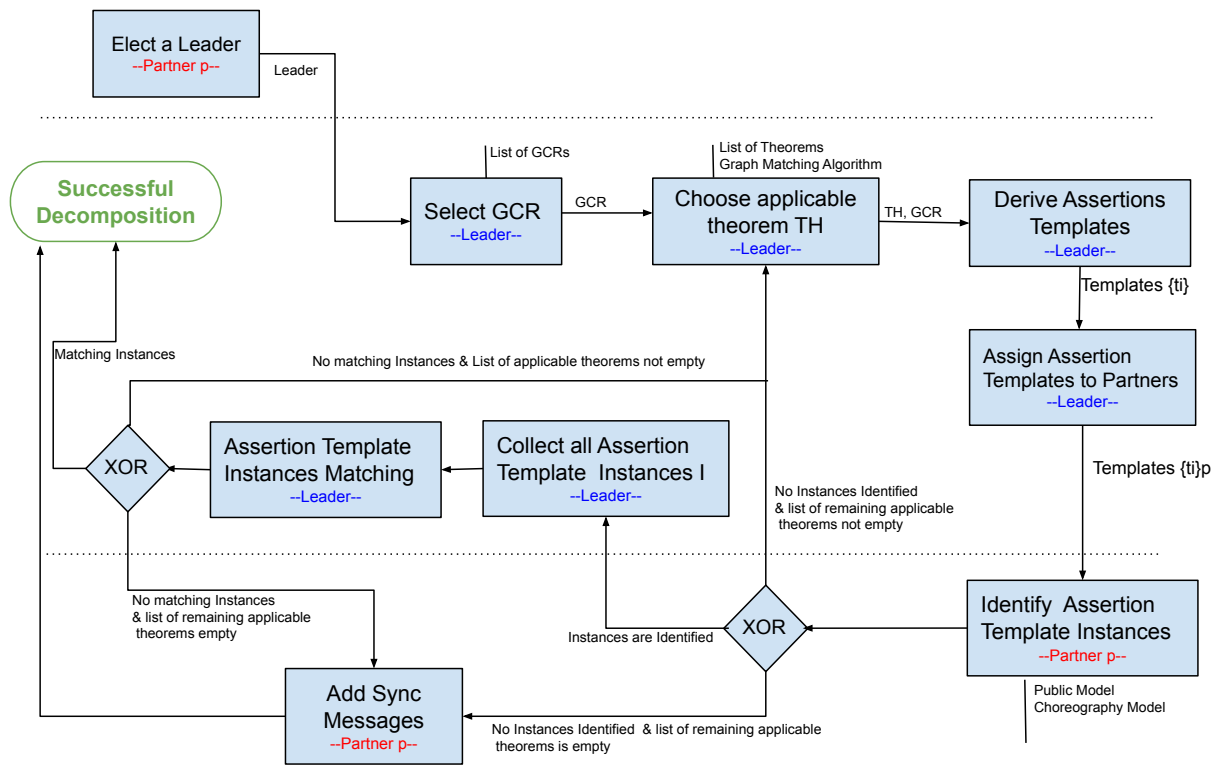

**Figure 10:** Decomposition Process

Given a GCR, the decomposition process starts by selecting a leader among the partners involved in the GCR. The leader will be responsible for identifying the pattern corresponding to the GCR (e.g., rightward chaining pattern or between pattern). This is trivial and can be accomplished with a simple exact graph matching algorithm (e.g., comparing node types and connectors). Once the pattern is identified, several decomposition theorems may be applicable. For example, in the case of the between pattern, Theorems 5, 6, and possibly 7 (if the processes are loop free) may be applied. The leader will then select and apply a theorem, derive the assertion templates accordingly, and send each of them to the corresponding partner–this step can be easily automated. An assertion template, in turn, is a derived assertion output from the theorem, where the actual interactions (i.e., message exchanges) have not yet been defined. For example, for $Middleman$, $\boxed{get\_permission\_of\_authority} \dashrightarrow \bigcirc{M}$ (where M shall be an activity interacting with $Special\_carrier$) is a derived assertion template of the global compliance rule $\boxed{get\_permission\_of\_authority} \dashrightarrow \bigcirc{safety\_check}$, as $M$ has not been defined yet.

Next, each partner will try to identify an assertion template instance that conforms with the derived assertion template–an instance of an assertion template corresponds to a template with actual activities. *Order_special_transport* is a valid option in this example as it interacts with *Special_carrier* fulfilling the template constraint (i.e., the template instance corresponds to $\boxed{get\_permission\_of\_authority} \dashrightarrow \text{(order\_special\_transport)}$). This can be automated by having each partner responsible for an assertion template iterating over the message exchanges in its private process model. Depending on the assertion template structure, one or several message exchanges may be selected for constructing an assertion instance candidate. The latter must conform with the assertion template. Moreover, it needs to be ensured that the process in question is compliable with it. Existing design-time compliance checking techniques can be employed in this regard (3; 4; 35; 6).

Several assertion instance candidates can be identified for the same partner, which may increase the probability of finding a collective solution among all partners. Afterwards, partners either collectively enter a negotiation phase and exchange their assertion instance candidates, or rely on the leader to collect all proposals and run a matching algorithm that selects assertion candidates, which replicate the templates derived by the decomposition. Indeed, two assertion templates may require that they use the same message exchange. Therefore, the matching algorithm will select the assertion instance candidates of different partners that have the same message exchange in common (using function $\psi$ or $\phi$ to ensure that the mapping is correct). While having the leader collecting the assertion proposals and doing the matching can be more efficient and reduce the communication overhead between partners, conducting the negotiation in a distributed manner reduces trust assumptions.

Note that it is possible to run the entire process in a distributed manner, without need for a leader. In this case, all partners will have to run the matching algorithms for identifying the GCR pattern. Moreover, they have to agree on the decomposition theorem to be applied (e.g., using a majority vote) and collectively execute the matching of assertion instances. If a matching solution is found, each partner will use the selected assertion instance for future run-time checking. Unless a solution is found, the next applicable theorem will be explored in the same way. If no solution could be found after trying all applicable decomposition theorems, synchronization messages become necessary for enabling distributed run-time checking of the GCRs. At run-time, no additional communication with other partners is needed for checking the GCR, unless a violation occurs. Similar to assertion and local compliance rules monitoring, each partner is responsible for complying with the derived assertions. This can be enforced using post-auditing processes by the respective legal entities, e.g., external audits conducted by data protection officers on hospitals participating in collaborative study on COVID-19 vaccines' efficiency. Indeed, in the healthcare sector, new methods exist, where it is possible to conduct a research study using federated machine learning[6]. In this setting, the ML application is conducted locally within each healthcare data provider infrastructure (e,g., hospitals or bio banks), and the resulting output models are aggregated instead. This prevents data of different participants from being merged in a central repository, and thus be subject to different and complex regulatory issues. These locally executed processes, nevertheless, also need to comply with data protection rules, where for example, the data used for the aggregated model must have patient consents beforehand. Therefore, external audits are necessary, at each site, to check whether each federated model used for the aggregation is indeed compliant with the GDPR requirements for data consents. Note that this also prevents collecting all participants' consents in one central repository for the purpose of compliance checks on the aggregated model.

Algorithm 1 realizes GCR decomposition as set out in Definition 2. It assumes that each node of the GCR is assigned to one partner being responsible for it. Further on, we assume the input GCR to be consistent and satisfiable (for dealing with unsatisfiable and inconsistent rules we refer interested readers to (11)). In the following, we first explain Algorithm 1 step by step and then illustrate the entire algorithm along Example 1 (see below).

Starting from the $\boxed{A}$ node (cf. Def. 2), Algorithm 1 walks outwards through all other nodes of the GCR. Thereby, the visited parts are copied and become assertions. Wherever the algorithm walks over a connector between two nodes $n$ and $s$, which are assigned to different partners $\rho(n)$ and $\rho(s)$, the GCR is split at this position as this dependency cannot be evaluated by a single partner. Next, the algorithm tries to replicate the connector where the GCR was split through (transitive) message exchanges between both affected partners by applying the transitive relationships from Section 2. Therefore, sets $n\bullet$, $\bullet s$, and $\Theta$ are calculated. Depending on the pattern of $s$ (cf. Def. 2), $n\bullet$ and $\bullet s$ contain the messages succeeding or preceding $n$ and $s$, respectively. Note that these calculations have to be accomplished in a decentralized manner by $\rho(n)$ and $\rho(s)$ themselves as $n$ and $s$ may be private tasks. Next, $\Theta$ combines those messages of $n\bullet$ and $\bullet s$ that can be combined.

---

[6]https://featurecloud.eu/

**Algorithm 1:** GCR decomposition DECOMPOSE(gcr)

- Global compliance rule $gcr = (N, \rho, \varphi, type, pattern)$
- Choreography model $y$, and $\mathcal{M}$ as the set of all partners' message nodes.
- We assume that $\rho$ also returns the partner private model of a node $n$.

select the only $a \in N$ with $pattern(a) = \boxed{A}$
initialize queue $Q \leftarrow \{a\}$
create (incomplete) Assertion $A_a \leftarrow \varepsilon\boxed{a}\varepsilon$ for the partner associated with $\rho(a)$
**foreach** $(n \leftarrow removeHead(Q))$ **do**
  **foreach** $(s \in N$ with $\varphi(n, s) \neq \emptyset)$ **do**
    $Q \leftarrow Q \cup \{s\}$
    **if** $(\rho(n) = \rho(s))$ **then**
      *//n and s involve the same partner*
      initialize $A_s \leftarrow @A_n$ as reference on $A_n$
      **if** $(pattern(s) = \boxtimes)$ **then** extend $A_s$ with $\varepsilon \dashrightarrow \boxtimes \varepsilon$
      **if** $(pattern(s) = \bigcirc{C})$ **then** extend $A_s$ with $\varepsilon \rightarrow \bigcirc{s} \varepsilon$
      **if** $(pattern(s) = \otimes)$ **then** extend $A_s$ with $\varepsilon \dashrightarrow \otimes \varepsilon$
    **else**
      *//n and s involve different partners $p_i, p_j$*
      **if** $(pattern(s) = \bigcirc{C})$ **then**
        $n\bullet \leftarrow \{m \in \rho(n) | m \in \mathcal{M}, \rho(n) \vDash \boxed{n} \dashrightarrow \bigcirc{m}\}$
        $\bullet s \leftarrow \{m \in \rho(s) | m \in \mathcal{M}, \rho(s) \vDash \boxed{m} \rightarrow \bigcirc{s}\}$
        $\Theta \leftarrow \{(m_n, m_s) \in (n\bullet \times \bullet s) | \gamma \vDash \boxed{m_n} \dashrightarrow \bigcirc{m_s}\}$

      **if** $(pattern(s) = \otimes)$ **then**
        $n\bullet \leftarrow \{m \in \rho(n) | m \in \mathcal{M}, \rho(n) \vDash \bigcirc{m} \dashrightarrow \boxed{n}\}$
        $\bullet s \leftarrow \{m \in \rho(n) | m \in \mathcal{M}, \rho(n) \vDash \boxed{m} \dashrightarrow \otimes\}$
        $\Theta \leftarrow \{(m_n, m_s) \in (n\bullet \times \bullet s) | \gamma \vDash \bigcirc{m_s} \dashrightarrow \boxed{m_n}\}$

      **if** $(pattern(s) = \boxtimes)$ **then**
        $n\bullet \leftarrow \{m \in \rho(n) | m \in \mathcal{M}, \rho(n) \vDash \boxed{n} \dashrightarrow \boxed{m}\}$
        $\bullet s \leftarrow \{m \in \rho(s) | m \in \mathcal{M}, \rho(s) \nvDash \boxed{m} \dashrightarrow \otimes\}$
        $\Theta \leftarrow \{(m_n, m_s) \in (n\bullet \times \bullet s) | \gamma \vDash \boxed{m_n} \dashrightarrow \boxed{m_s}\}$

      **if** $(\Theta \cup (n\bullet \cap \bullet s) = \{\emptyset\})$ **then**
        *//No implicit dependency between n and s*
        add *sync* message between $n$ and $s$
        update models $p_1, \ldots, p_n$, and $\gamma$
        recalculate $n\bullet, \bullet s$, and $\Theta$
      **else**
        *//implicit dependency m between n and s exists*
        select $(m_n, m_s) \in \Theta \cup (n\bullet \cap \bullet s)^2$
        **if** $(pattern(s) = \bigcirc{C})$ **then**
          extend $A_n$ with $\varepsilon \dashrightarrow \bigcirc{m_n} \varepsilon$
          create Assertion $A_s \leftarrow \varepsilon\boxed{m_s} \rightarrow \bigcirc{s} \varepsilon$ for $\rho(s)$
        **if** $(pattern(s) = \otimes)$ **then**
          extend $A_s$ with $\varepsilon \bigcirc{m_n} \dashrightarrow \varepsilon$
          create Assertion $A_s \leftarrow \varepsilon\boxed{m_s} \dashrightarrow \otimes \varepsilon$ for $\rho(s)$
        **if** $(pattern(s) = \boxtimes)$ **then**
          create Assertion $A_s \leftarrow \varepsilon\boxed{m_s} \dashrightarrow \boxtimes \varepsilon$ for $\rho(s)$

  **foreach** $((s \in N$ with $\varphi(n, s) \neq \emptyset))$ **do**
    *//same as for each $(n, s) \in C$ above*
    *//but with flipped directions*

  **foreach** (*partner i*) **do**
    **foreach** $((A_j, A_k)$ *of partner i*) **do**
      **if** $(A_j, A_k$ *have the same* $\boxed{A}$ *pattern*) **then**
        merge $A_j$ and $A_k$ based on the $\boxed{A}$ pattern

  **foreach** (*Assertion A*) **do**
    **if** (*A has empty* $\bigcirc{C}$ *and* $\otimes$ *patterns*) **then** remove $A$

If *s* is a Ⓒ node (i.e., *s* must follow *n*), Θ contains message tuples $(m_1, m_2)$ that ensure that *n* is always followed by $m_1$, $m_1$ by $m_2$ (unless $m_1 = m_2$), and $m_2$ by *s*. Any pair $(m_1, m_2) \in \Theta$ can then be used to complement the created assertions, i.e., $m_1$ becomes a placeholder for *s* within the assertion of $\rho(n)$, whereas $m_2$ replaces *n* for $\rho(s)$.

Regarding ⊗ nodes (i.e., *s* must not follow *n*), all pairs of messages $(m_1, m_2) \in \Theta$ ensure that *n* is preceded by $m_1$ in any case, and $m_1$ is preceded by $m_2$ (unless $m_1 = m_2$), whereas *s* never follows $m_2$. Finally, for ⋈ nodes an occurrence of *s* after *n* allows ignoring the rule. Hence, ⋈ nodes result in pairs $(m_1, m_2)$ such that $m_1$ may only occur after *n* and $m_2$ may only occur after $m_1$ (unless $m_1 = m_2$). However, there should be at least one case in which $m_2$ is followed by *s* (i.e., *s* is not always preceded by $m_2$).

Finally, all assertions of the same partner, which depend on the same Ⓐ message, are merged to reduce the number of assertions. Remaining assertions without consequences are removed as they result from the processing of ⋈ nodes, but have not been merged in the previous step. Remember that ignoring ⋈ nodes is allowed as this makes rules even more strict.

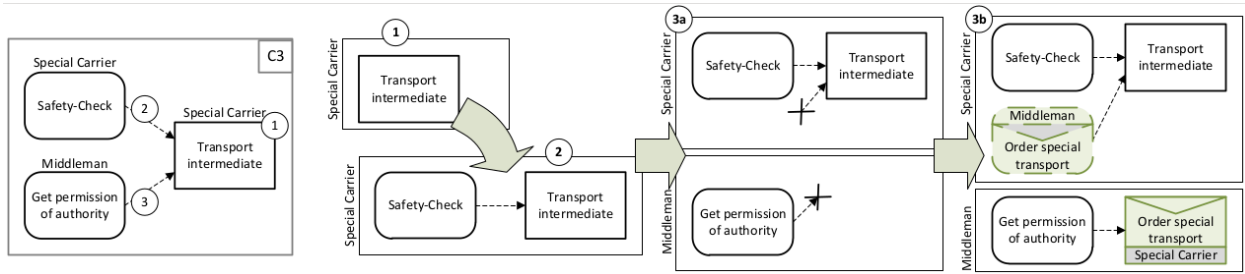

**Figure 11:** Decomposition process of GCR *C*3

**Example 1.** *Let us apply Algorithm 1 to GCR C3 (cf. Fig. 11) from the running example that we introduced in Section 1 (cf. Figs. 1 - 4). Let the responsibilities be ρ(Safety Check) = Special Carrier, ρ(Get permission of authority) = Middleman, and ρ(Transport intermediate) = Special Carrier. After assigning responsibilities, Algorithm 1 starts with ⒶⒶ node* Transport intermediate *and creates a new assertion for the* Special Carrier *who is responsible for this activity. Then the* Safety check *is discovered and added to the assertion, since it belongs to the same partner. In turn, another partner (i.e.,* Middleman*) is responsible for activity* Get permission of authority*. Hence, the algorithm cuts the respective connector and creates a new assertion for the respective partner. Next,* Special Carrier *and* Middleman *determine n• and •s with n• = {Waybill, T. Details, Req. Details, Order ST} and •s = {Order ST} to calculate those message pairs* Θ = {(Waybill, Order ST), ..., (Order ST, Order ST)} *that can be used to transitively replicate the connector where the GCR was split. Finally, the algorithm places the selected messages into both assertions in such a way that the correctness of the original rule is preserved through the (leftwards) transitivity of eCRGs. Note that the* Special Carrier *could use message* Waybill *instead as* (Waybill, Order ST) ∈ Θ *holds.*

For the same GCR, it is possible to infer several decomposition alternatives, depending on which interactions are used to find a transitive control flow relationship between the nodes of the GCR. It is also possible that no direct link can be identified between two partners' GCR nodes (i.e., there is no interaction between these two partners). As such, interactions with *intermediary partners* can be used to find an indirect link (i.e., transitive interactions). As aforementioned, if no transitivity is identified between the GCR nodes of two partners (even not through intermediary partners), it becomes necessary to exchange additional execution data between the partners involved in the GCR, by, for example, adding sync messages. Sync messages are a specific type of messages communicated between partners to inform about the state of a given task (e.g., terminated, started, not executed). Although sync messages are not preferred as they expose private data about the exact execution time of a private task, they become necessary when the GCR cannot be decomposed into assertions, i.e., no transitive relations can be identified.

In the following, we discuss the complexity of the GCR decomposition in Algorithm 1. Results on checking regulatory compliance in general have been provided in (59). The first and second loops iterate over the nodes of the compliance rule. If we consider that two nodes can only have one flow connector, the number of required operations will be $n\frac{n-1}{2}$, otherwise $n(n-1)$. In both cases complexity corresponds to $O(n^2)$. The first *if* statement is $O(1)$, whereas the *else* statement calculates $n\bullet$, $\bullet s$ and $\theta$ each with a worst case complexity of $O(n^2)$. The second inner loop has the same complexity as the first inner loop. The third nested inner loop iterates over partners and compare assertions within the same partner with a number of operations equal to $n \times m \frac{m-1}{2}$, which gives a complexity of $O(n^3)$. Finally, the last inner loop has a complexity of $O(n)$. Obviously, the overall worst case complexity of the algorithm is polynomial $O(n^4)$; i.e., outer loop combined with the third nested inner loop.

## 5. Verifying GCR Decomposition

The GCR decomposition algorithms from Section 4 are based on the theorems we have presented and proven in Section 3. Although these theorems support most control flow (i.e., behavior) compliance patterns known from literature (56; 19), they cannot cover every possible scenario. There may be two reasons for this: (i) either the structure of a GCR is not covered by Theorems 1–8 or (ii) none of the proposed decompositions is applicable. In both cases, it might not only become necessary to find a novel decomposition of a GCR, but also to verify the latter, i.e., to prove the correctness of the decomposition. One approach to accomplish this would be to apply the eCRG semantics and to formally prove correctness (cf. Section 3). However, this is far from being trivial. Therefore, we introduce Algorithm 2 that enables the automated verification of GCR decompositions based on eCRG model checking.

---

**Algorithm 2:** Verification of Decompositions (Assertions $A_1, \ldots, A_n$, GCR $gcr$)

1 Let $\mathcal{A}$ be a function that translates an eCRG into a corresponding finite-state automaton
2 $\mathcal{M} \leftarrow \bigcap_{1 \leq i \leq n} \mathcal{A}(A_i)$
3 $\mathcal{R} \leftarrow \mathcal{M} \cap \neg \mathcal{A}(gcr)$
4 **if** $\mathcal{R} = \emptyset$ **then**
5    *//Decomposition is correct*
6    **return** true
7 **else**
8    *//Decomposition is incorrect*
9    **return** any arbitrary trace through $\mathcal{R}$ as counter example.

---

The main idea of Algorithm 2 is to interpret a GCR decomposition as declarative process model and to verify whether it solely allows for execution traces that comply with the original GCR. Thus, techniques that are known from the verification of declarative process models (50; 52) can be applied: First, all assertions $A_1, \ldots, A_n$ of the decomposition are translated into finite state automatons $\mathcal{A}(A_1), \ldots, \mathcal{A}(A_n)$. Their intersection $\left(\bigcap_{1 \leq i \leq n} \mathcal{A}(A_i)\right)$ corresponds to an automaton that only accepts traces that comply with every assertion. In turn, $\neg \mathcal{A}(gcr)$ denotes the automaton that accepts solely traces violating the original GCR. If the intersection of these two automatons is empty, the decomposition is correct as it only allows for traces that do not violate the original GCR and, thus, comply with it.

$$\left(\bigcap_{1 \leq i \leq n} \mathcal{A}(A_i)\right) \cap \neg \mathcal{A}(gcr) = \emptyset \Rightarrow \left(\bigcap_{1 \leq i \leq n} \mathcal{A}(A_i)\right) \subseteq \mathcal{A}(gcr)$$

For any choreography y, whose partners ensure $A_1, \ldots, A_n$, we can now directly conclude:

$$\mathcal{A}(y) \subseteq \left(\bigcap_{1 \leq i \leq n} \mathcal{A}(A_i)\right) \Rightarrow \mathcal{A}(y) \subseteq \mathcal{A}(gcr), \text{i.e., } y \text{ complies with } gcr$$

## 6. Implementation

The presented approach is implemented as part of the C³Pro framework[7], which deals with change and compliance in process choreographies (21). The framework provides sophisticated functions for defining, propagating and negotiating changes in the context of process choreographies. Furthermore, it comprises a modeling component as one of its core components for editing and changing public and private process models as well as process choreography models. This component further enables the visualisation of change propagations. In the context of the present work, the three-layer architecture of the framework (21) (i.e., process modeling, change, and execution) was **extended with** additional components for dealing with **process compliance**. In detail, these new components include (i) an eCRG modelling tool, (ii) an automated generator of compliant choreographies, (iii) a model checker, and (iv) a GCR decomposition tool.

---

[7]http://www.wst.univie.ac.at/communities/c3pro/

Figure 12 depicts the main components of the C$^3$Pro framework. The **compliance (CME) and process modeling (PME) environments** allow defining and editing *compliable* process choreography models (39; 35) and decomposing global compliance rules. Compliability was introduced as *"a semantic correctness criterion to be considered when designing interaction models. It ensures that interaction models do not conflict with the set of imposed global compliance rules"* (39). At design time, it is ensured that the created choreography models are compliant with the defined compliance rules. Using both PME and CME, it becomes possible to parameterize and automatically generate compliable choreographies, which then can be used for testing and simulation purposes. A user, therefore, can specify the number and patterns of the GCRs as well as the parameters of the process and choreography models (e.g., number of private and public tasks, number of partners and interactions, or number and types of the control flow patterns that shall be covered by the processes) (6). Although the generated models represent synthetic processes without real-world semantics (i.e., these models do not reflect actual use cases such as a manufacturing collaborative process), they may serve as a support for simulation and research work evaluation, e.g., model executions can result in distributed logs of synthetic data, which are useful for evaluating the efficiency of specific mining techniques. In the context of this work, this tool can be used to test the feasibility and applicability of the decomposition process on more complex choreographies and corresponding GCRs. Currently, the automated generator tool only supports four basic compliance patterns. However, other GCR patterns can be directly inferred from the models and be used for testing. The tool is integrated in the C$^3$Pro framework and can be tested. A data set of automatically generated models and the corresponding GCRs are made available in the same repository.[7] Finally, the change editor allows defining and editing changes of process models and compliance rules, respectively.

The **Compliance Management Service** represents the *main extension* to C3Pro related to this work, and handles the defined compliance rules and implements the theorems as well as the GCR decomposition algorithm (cf. Section 4). As process execution engine, the Cloud Process Execution Engine[8] is utilized. Most functions of the C$^3$Pro framework are provided as a RESTful service, which enables unified access from any client being able to communicate via HTTP. Finally, the Compliance Management Service serves as a pluggable middleware that may be used to integrate other process execution engines.

For testing the framework, we edited BPMN 2.0 choreography and collaboration models using Signavio[9] and exported them to the C$^3$Pro framework as XML files. Examples are extended with GCRs, which are then decomposed into derived assertions using Algorithm 1. To this endeavor, mainly the CME, PME and the compliance management service were used.

In addition, the C$^3$Pro framework was *extended with* a novel **eCRG model checker** that was published on github[10]. Its command line interface enables the specification and verification of both global and local compliance rules (GCR and LCR) as well as process models and choreographies. In particular, the eCRG model checker supports the verification of

- GCR decompositions, i.e., it allows verifying whether GCRs can be concluded from a given decomposition,

- local compliance, i.e., it allows verifying whether a single process model complies with a given compliance rule (CR), and

- global compliance, i.e., it allows verifying whether a process choreography complies with a given GCR.

Moreover, the eCRG checker enables the automated decomposition of tree-structured GCRs with a single antecedence.

In order to verify GCR decompositions and the various kinds of compliance respectively, the eCRG model checker translates global and local complicance rules as well as process models and choreographies into automaton, which are then combined and intersected. Depending on whether the resulting automaton is empty or not, the verification has been successful or a counter example trace has been produced. The eCRG model checker has been written in Java 8 on top of the `dk.brics.automaton` framework[11].

---

[8]http://cpee.org
[9]http://academic.signavio.com/
[10]https://github.com/davidknuplesch/SCV

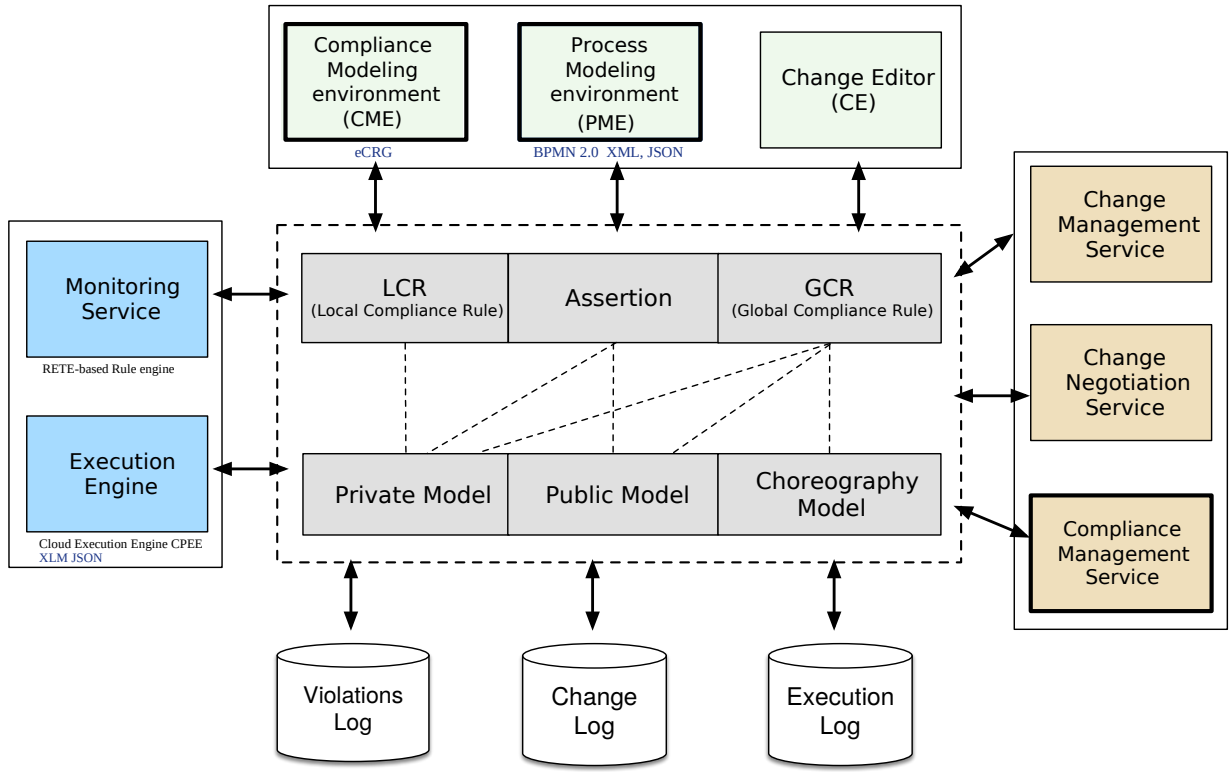

**Figure 12:** C3Pro Prototype Architecture

## 7. Applicability

This section demonstrates and discusses how the presented decomposition algorithm can be applied in real-world settings. Here, a GCR may be imposed on process choreographies by *external* sources, e.g., considering regulatory documents such as the GDPR or standards such as ISO norms (63). A GCR may also reflect *internal* compliance rules expressing, for example, implicit dependencies between the partners that are crucial for (re-)scheduling the process activities for one partner or across multiple partners in the choreography. The visibility of activities and compliance rules in real-world settings depends on the contractual situation and the roles of the partners in the choreography. In supply chains in automotive domain, for example, an Original Equipment Manufacturer (OEM) might demand insights into certain specifics of the private processes of its suppliers and the connected (internal) compliance rules.

### 7.1. Use Cases

In the supply chain example presented in the Section 1, GCR $C3$ (cf. Fig. 2) reflects an externally imposed GCR on safety in manufacturing and logistics processes. GCR $C1$, in turn, might reflect an internal quality assurance rule that is solely verified by the *Manufacturer*, but is also made visible to the other partners in order to, e.g., create trust.
The real-world use case from manufacturing depicted in Fig. 13 demonstrates how the decomposition algorithms can be employed to lift implicit (data) connections to explicitly modeled assertions. The use case covers a process choreography between `Partner 1` (i.e., a car manufacturer), `Partner 2` (doing injection molding), and `Partner 3` (i.e., the electro plater that coats parts in a correct color scheme). The choreography is designed and implemented using the CPEE (Cloud Process Execution Engine)[12]. Figure 13 shows the public activities of all partners, e.g., activity `place` order for `Partner 1` and private activities, e.g., activity `wait for order completion` for `Partner 1`. The public activities send or receive messages, e.g., activity `place order` sends a message received by activity `receive order`. Note that the scenario abstracts from the details of the public and private activities, which are modelled as sub-processes activities, except for the `electroplating` task where the corresponding sub-process is depicted. The sub-process describes the measuring of the quality of the bath and the glossiness; both measures are then forwarded to the partner. Depending on the measures bath maintenance is conducted (alternative default branch). All sub-processes are of different complexity, i.e., they might contain decisions and loops, as well. The complex activity `wait for order completion`, in particular, comprises a set of sub-activities and is signifying the scheduling between the activities of the different partners.

---

[11] https://www.brics.dk/automaton/
[12] https://cpee.org/

During the design of the choreography the partners specified implicit connections, i.e., dependencies between (private) activities that are not covered by message exchanges and express mostly data dependencies. For example, activity `wait for order completion` (private, `Partner 1`) implicitly depends on the the data produced by activities `prepare for manufacturing`, `manufacturing of parts`, and `quality control` (all private, `Partner 2`).

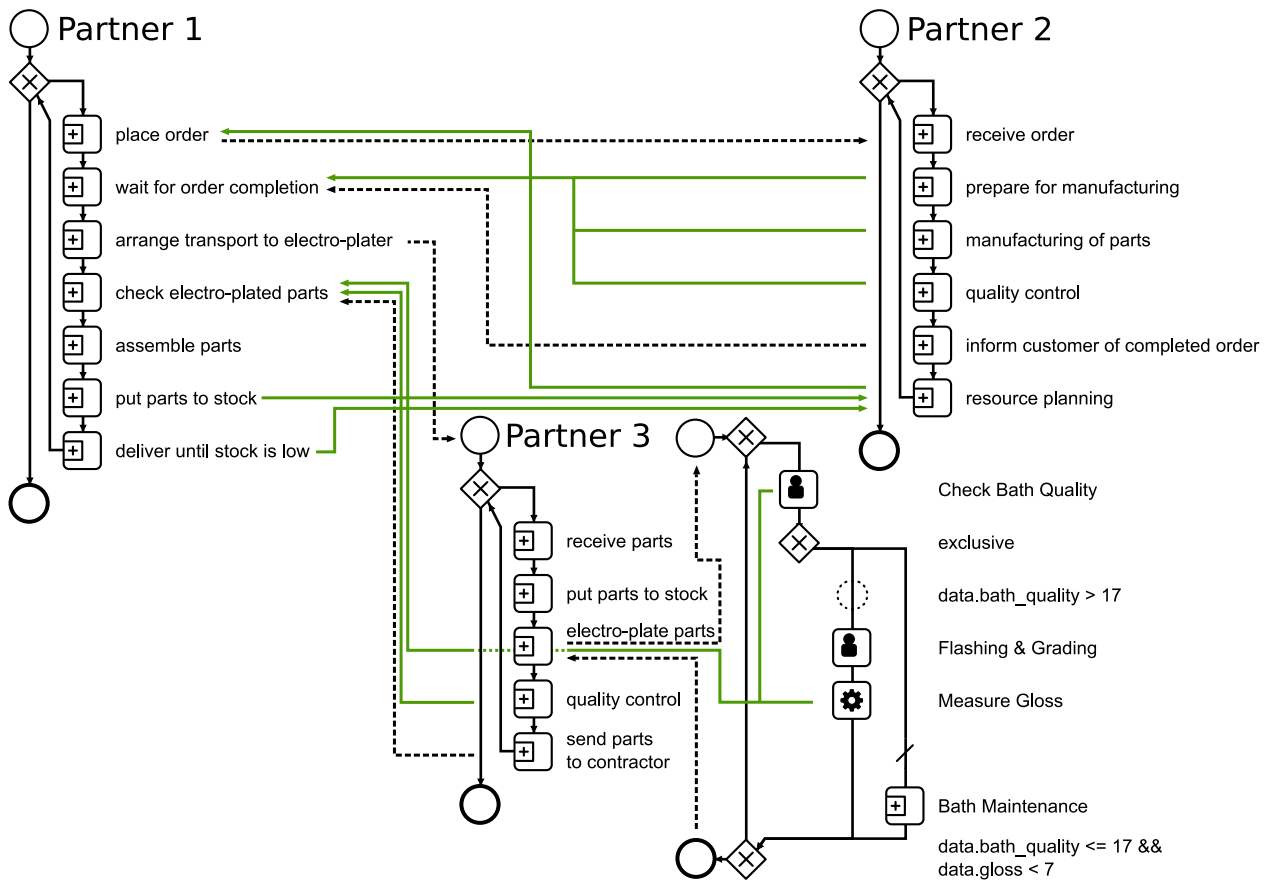

**Figure 13:** Manufacturing Use Case

These implicit connections refer to two main "functions" of the manufacturing setting, i.e., (i) *resource planning* and (ii) *quality control*.

(i) Proper resource planning is part of the contract between the partners. There are implicit rules regarding how fast `Partner 2` has to react to an order. This depends on assumptions how fast the stock drops for `Partner 1`. This manifests as follows: If activity `place order` (`Partner 1`) occurs, then activity `resource planning` `Partner 2` has been done before (i.e., resource planning data received) and activity `resource planning` (`Partner 2`) was based on activities `put parts to stock` and `deliver until stock is low` (`Partner 1`). Understanding this as a compliance task, we can say that when the above activity information matches, the compliant ordering can be ensured.

The question is how such implicit connections can be checked without revealing information on the private activities. Here, the presented decomposition algorithm can help. The idea is to express the implicit connections by GCR and verify them based on assertions. Take, for example, the implicit connection between private activities `place order` (`Partner 1`) and `resource planning` (`Partner 2`). We can formulate this implicit connection as GCR
$C_1$ : `place order` ⇢ `resource planning`
Using Algorithm 1, $C_1$ can be decomposed into the following assertions:

- $A^1_{C_1}$ : `place order` ⇢ $m_1$

- $A^2_{C_1}$ : $m_1$ ⇢ `resource planning`

Note that doing so, the implicit connection is lifted up to an explicit one by sending message $m_1$.

(ii) Consider now the more complex GCR covering the overall `quality control` that involves all three partners. `Partner 1` has to do the final inspection, `Partner 2` has to ensure the quality of the injection molded parts (no cracks, no holes), and `Partner 3` has to ensure the quality.
It is assumed that data for checking quality individually has been delivered whenever an order is finished. In this case, activity `wait for order completion` by `Partner 1` yields all information about the quality of the injection molded parts and the electroplating process, whereas during activity `check electro-plated parts` by `Partner 1` all additional information about the molding process is available. Overall, if activity `check electro-plated parts` (`Partner 1`) occurs, then it has to be checked whether or not activities `electro-plate parts` and `quality control` by `Partner 3` were executed. Before that, for activity `wait for order completion`, activities `prepare for manufacturing` (e.g. machine calibration data, material information), `manufacturing of parts`, and `quality control` by `Partner 2` were executed. If all of the above information has been received, the manufacturing process was compliant, i.e., all required steps seem to have taken place.

## 7.2. Discussion
In the introduction, the following research question was stated:

> *RQ: How to verify GCRs in a decentralized setting of a process choreography where no central coordinator with complete knowledge on the private and public tasks of all partners exists?*

In the following, we discuss how far the article at hand has addressed RQ and which open questions still remain. For this purpose, we sketch the end-to-end application of the decomposition algorithm along the following steps:

1. Check whether the GCR can be verified at choreography level, i.e., solely referring to interactions.
2. Check whether the GCR can be verified at the public process of one partner.
3. Check whether the GCR can be verified on the public processes of at least two partners: partners have to check for the absence/presence of GCR-related activities and activity orders based on choreography and interactions if possible. Otherwise, verification has to be postponed to the runtime by additionally synchronizing activities OR compliance has to be verified in an ex post way based on logs if available.
4. Otherwise: The GCR refers to the private processes of one or several partners and a decomposition has to be applied.

As private activities and their dependencies are not visible to all collaborators, parts of the decomposition algorithm are executed locally by all partners involved in this GCR in order to identify possible transitive relations between their corresponding private activities and possible public activities, or interactions that replicate the connector where the GCR was split. This results in multiple derived assertion alternatives, which are then aggregated to alternatives from other partners in order to find a combination that recreates the original rule as described in Section 4 (cf. Example 1). Once the GCR is decomposed and the corresponding assertions are derived, each partner locally checks its derived assertions at runtime.

Overall, RQ has been addressed in breaking down the problem of GCR verification on distributed processes into the steps outlined above. Moreover, a sophisticated decomposition algorithm for GCR that refers to private activities of one or multiple partners has been provided. This enables distributed compliance verification for the supply chain and manufacturing use cases discussed in Section 7.1.

**Limitations and open questions:** This paper focuses on structural compliance, i.e., a GCR solely refers to control flow patterns. Compliance patterns that deal with, for example, data and resources (cf. (61)) are future work and will add substantially to the complexity of the approach. Moreover, we have applied our approach to a use case from the manufacturing domain. However, additional studies in other domains, such as healthcare and logistics, become necessary to evaluate the generalizability and broad applicability of the approach. In addition, the presence of ***XOR*** branches in the processes (where sending of messages on these branches is optional) does not affect the correctness of the decomposition as long as the processes are ***compliable*** with the original GCR (39). As aforementioned, we assume the soundness of the different process models (i.e., consistency and compatibility) and their compliability to the original GCR. This means that original GCRs are correctly specified, and the decomposition enables their checking in a distributed way. In this case, transitivity ensures correct decomposition of GCR even at the presence of XOR branches. If no transitive relations are identified, sync messages are required. Further on, in the end-to-end approach outlined above, Step 3 still offers the challenge on how to check GCR on public processes of multiple partners.

## 8. Related Work

The work presented in this article can be positioned at the interface between process choreographies and business process compliance. Section 8.1 summarizes basic works from these two research fields, whereas Section 8.2 discusses approaches that address issues at the interface between them.

### 8.1. Basic Research Fields

Section 8.1.1 gives backgrounds on process choreography research, whereas Section 8.1.2 summarizes basic works dealing with business process compliance.

#### 8.1.1. Process Choreography

*Process choreography* research has mainly dealt with the modeling of process choreographies and the verification of correctness properties. For this purpose, specific choreography modeling languages like *Let's dance*, *Interaction Petri nets*, and *BPMN choreography diagrams* are proposed, which support the modeling of collaborative process behavior. A particular focus of existing works has been put on correctness properties of choreography models (e.g., realizability), which have been intensively studied in literature (17; 16)–for an overview we refer interested readers to (62).

#### 8.1.2. Business Process Compliance

*Business process compliance*, in turn, has been investigated for more than a decade and several surveys exist (e.g., (4; 24)). Contemporary approaches have focused on compliance rule languages, including visual notations (3; 40; 38), logic-based formalisms (45; 43), and Event Calculus (49). Moreover, several approaches enable compliance checking at both build- and run-time (e.g., (4; 44; 40)) or cover different process perspectives of compliance rule checking such as behavior, data, time, and resources (58; 40). Finally, characteristic patterns for business process compliance are proposed by (19).

A formal approach that verifies local process behavior (i.e., WS-BPEL process models) against legal constraints, specified in terms of the *Compliance Request Language*, is proposed by (19). This approach and similar works focus on local compliance rules, which can be checked for a given (partner) process model. By contrast, we consider verifying global compliance rules (GCR) in a process choreography based on their correct and lossless decomposition into assertions that can be checked locally by each concerned partner.

Another related field deals with checking compliance of a (local) process model against its *refinement* or *implementation*. An approach that enables checking compliance of a (local) process model against its refinement is presented in (53). More specifically, this approach deals with the automated verification of lower-level against higher-level UML activity charts. Behavioural containment is established to ensure that a lower-level chart constitutes a valid refinement of the higher-level one. Similarly, (18) presents an approach for enforcing compliance of hierarchical business processes with visually specified security constraints. An approach that enables checking compliance of a (local) process model against its implementation is presented in (8), which derives the specification of a web application from a (local) process model followed by the verification of web execution logs against derived LTL formulae. Although the problem addressed by these approaches is different from the one considered in our paper, the techniques could be of interest for global compliance checking in choreographies as well.

**8.2. Interface between Process Choreography and Process Compliance**

There exist several approaches that address issues at the interface between process choreography and process compliance. Section 8.2.1 discusses centralized and distributed approaches for checking compliance in multi-party processes (i.e., process choreographies). In turn, approaches that map global contracts (i.e., sets of global compliance rules) to compliable process choreographies are presented in Section 8.2.2. The conformance between process choreography and local partner processes are considered in Section 8.2.3. Finally, issues related to business process compliance in the context of dynamically evolving business partner networks are discussed in Section 8.2.4.

*8.2.1. Compliance Checking in a Process Choreography – Centralized vs. Distributed Approaches*

Compliance checking mechanisms assuming a trusted party are proposed by (27). In (26) the same authors present a service-oriented approach that relies on a central *integration platform* in order to enable cross-organizational service interactions, while at the same time meeting global compliance rules. As opposed to our work, this approach relies on a central component (i.e., the integration platform) to ensure that global compliance rules can be checked.

(33) advocates compliance checking in a distributed process (i.e., process choreography) as crucial, but it cannot be assessed in how far the approach deals with the restricted visibility and availability of process information as we do. In prior work, we have introduced the criterion of compliability (39) that captures the ability of a choreography to comply with a given set of compliance rules at all and how to check it (35). The approach presented in (37) enables checking the effects of changes on the compliance in process choreographies based on dependency graphs between global and local compliance rules as well as assertions. Finally, (22) provides an overview on the challenges, related approaches, and possible solutions at the interplay of compliance, change, and choreographies.

Distributed approaches that rely on IoT technology are proposed by (51; 48). The approach suggested by (51) considers the flow of physical objects between the parties of a multi-party process. In particular, this approach exploits the sensing capabilities of smart devices to improve process compliance checking. For this purpose, commitments define mutual contractual relationships between parties in a declarative way and drive the configuration of smart devices, which check their satisfaction and, in case of misalignment, inform the concerned parties timely.

This multi-party process compliance monitoring approach is conceptually enhanced by (48) through IoT-enabled artifacts. This approach proposes a decentralized solution switching from control- to artifact-based monitoring, where physical objects can monitor their own conditions as well as the process activities in which they participate, i.e., compliance monitoring is distributed among the physical artifacts interacting with the global process. To instruct these smart objects, BPMN process models are translated into a set of artifact-centric process models, rendered in Extended-Guard-Stage-Milestone (E-GSM) notation. In particular, this work shows that artifact-based modeling approaches have a high potential in respect to multi-party process management involving physical objects, which has not been the focus of our work.

Finally, (46) discusses legal, organizational, human-centered, technical and economic challenges to be tackled in the field of business process compliance when enacting the (cross-organizational) business processes on the Ethereum blockchain. For example, at the implementation level, the immutability of illegal content or the error-proneness and zero-defect tolerance of smart contracts raise challenging issues in this context. Although this work does not deal with a concrete compliance verification approach, it indicates directions for future research on process compliance when using blockchain infrastructures for enacting multi-party business processes.

*8.2.2. Mapping Global Contracts to Process Choreographies*

Contract languages allow specifying obligations, permissions and prohibitions in business contracts, which govern the interactions between business partners. (28) provides means to model corresponding contract constraints. An early approach that extends choreographies with such contract constraints is provided by (5). This approach transforms the contract constraints into expressions of a choreography language, i.e., contract terms are translated into choreography expressions that govern the global process (i.e., choreography) to ensure compliance. In particular, it is shown how cross-organizational business processes can be monitored and enforced according to business contract specifications through the transformation of the contract definition to constraints on (global) process behavior. However, this approach is less powerful than ours as it tightly couples compliance constraints with choreography models, which aggravates the evolution of both choreography model and contract constraints. Besides, this approach does not consider local compliance checking (i.e., locally checking assertions derived from the decomposition of global compliance rules), which limits its applicability at the presence of more complex compliance requirements.

(31) advocates Dynamic Condition Response (DCR) Graphs for decomposing global contracts into local processes. More precisely, (31) shows how a timed DCR Graph can be used to describe the global contract for a timed multi-party process (i.e., choreography), which can then be distributed as a network of communicating timed DCR Graphs (i.e., local processes) describing the local contract for each party. As opposed to our work, this approach relies on a declarative process modeling approach with a focus on discrete time deadlines.

### 8.2.3. Conformance between Process Choreography and Local Partner Processes

Several proposals have been made to ensure *conformance* between choreography (i.e., the global process) and the local processes of the involved business partners. In (1), conformance checking of the event logs of local processes against a given choreography model is addressed. As such an event log is not available at design time, (43) relies on a graph transformation tool–GROOVE (GRaphs for Object-Oriented VErification)–to automatically verify that a local process of a partner involved in a choreography complies with the globally specified behavior of that choreography. LTL semantics of the choreography model is employed and token semantics of the local process model, which is expressed in terms of a BPMN collaboration diagram, is used to verify conformance.

(9) relates the theory of contracts with the notion of *choreography conformance*, i.e., it is checked whether an aggregation of services correctly behaves according to a high level specification of their possible conversations. For this purpose, a method based on the composition of choreography projection and contract refinement is presented that allows verifying that a service with a given contract can correctly play a specific role within a choreography. Finally, (14) presents an approach for ensuring conformance between a set of BPMN collaboration diagrams (i.e., local process models) and a BPMN choreography diagram (i.e., choreography model).

As opposed to these approaches, our decomposition-based method verifies the compliance of a choreography model with global compliance rules and regulations that cover multiple process perspectives. However, conformance between the choreography and the participating partner processes can be considered as a prerequisite of our approach.

### 8.2.4. Ensuring Compliance in Dynamically Evolving Partner Networks

(64) assumes that the partners in a business networks try to provide wrong information and, hence, introduce the notion of accountability. Compliance requirements also need to be met in dynamic business networks (12; 13). In such a network, the partners may join and leave the collaboration dynamically and tasks over which compliance rules may be specified then have to be distributed or delegated to new partners or be backsourced by network participants in order to avoid compliance issues. In (12; 13), a conceptual model for aligning the compliance requirements in a business network with the monitoring requirements they induce on the network participants, particularly when the network changes or evolves, is presented. Various techniques (e.g., task delegation and in-house backsourcing) for ensuring the consistency between the monitoring and compliance requirements as well as metrics for evaluating the status of a collaboration in respect to compliance monitorability are discussed. Obviously, this approach lacks a process perspective, but is complementary to our work with a focus on business network changes and their effects on compliance requirements.

## 9. Conclusions

The interplay between interoperability and business process compliance poses a tremendous challenge on companies. In this problem space, the work at hand addresses the question on how to verify global compliance rules (GCR), i.e., rules that span multiple partners in a multi-party process (i.e., process choreography), in a decentralized manner where certain tasks of one partner process might not be visible to the other partners due to confidentiality reasons. The presented approach focuses on the decomposition of a GCR such that the decomposed parts, i.e., assertions, can be checked by the partners locally. Consequently, compliance verification is shifted from a global to a decentralized manner.

The presented decomposition approach addresses research question *RQ*, which we introduced in Section 1. In particular, the presented decomposition algorithm exploits transitivity properties of the GCR for finding the correct decompositions. The correctness is formally proven. Moreover, the complexity of the decomposition algorithm is formally analyzed and also illustrated based on specific scenarios. The feasibility of the decomposition algorithm is shown based on a prototypical implementation, including a model checker for ensuring decomposition correctness. The applicability of the approach is demonstrated through a use case from the manufacturing domain.

Future work targets at two directions, (1) GCR language and (2) applications. First, we want to extend the decomposition based on the eCRG formalism, as used in this work, to arbitrary GCR and adapt the decomposition to other compliance rule languages such as Declare (55), PENELOPE (25), or BPMN-Q (3). Second, further applications of the approach include healthcare as interoperability and compliance are *"strategic imperatives"* in this domain.

Overall, the presented approach provides a fundamental brick in enabling process collaborations between different partners by infringing neither their privacy nor any regulations.

## Acknowledgment

This work is jointly supported by (i) the competence center SBA Research (SBA-K1) funded within the framework of COMET Competence Centers for Excellent Technologies by BMVIT, BMDW, and the federal state of Vienna, managed by the FFG, (ii) The FFG ICT of the Future project 874019 dIdentity & dApps, (iii) the "Austrian Competence Center for Digital Production" (CDP) under contract number 85418, and (iv) the European Union's Horizon 2020 research and innovation programme under grant agreement No 826078 (FeatureCloud). We thank Juergen Mangler for his input on the manufacturing use case.